\documentclass{aa}

\usepackage{txfonts}
\usepackage[T1]{fontenc}

\usepackage{graphicx}	
\usepackage{amsmath}	
\usepackage{amssymb}	
\usepackage{natbib}
\usepackage{multirow} 
\usepackage{bm} 
\usepackage{threeparttable}
\usepackage{hyperref}
\usepackage{epstopdf}
\usepackage{float}
\usepackage{mathrsfs}
\usepackage{array,multirow}
\usepackage{booktabs}
\usepackage{bm}
\usepackage{anyfontsize}
\usepackage{changepage}     
\usepackage{xcolor}
\usepackage{afterpage}
\usepackage{placeins}
\usepackage{caption}

\begin{document}

   \title{The Relationship Between Emission Line and Continuum Luminosity and the Baldwin Effect in Blazars. I. The Case of the Mg II $\lambda$2798 \AA\ Emission Line}

   \author{V\'ictor M. Pati\~no-\'Alvarez \inst{1,2} \fnmsep\thanks{Corresponding author: victorm.patinoa@gmail.com}
          \and
          Jonhatan U. Guerrero-Gonz\'alez \inst{1}
          \and
          Vahram Chavushyan \inst{1}
          \and
          Douglas E. Monjardin-Ward \inst{1}
          \and
          Tigran G. Arshakian \inst{3,4,5}
          \and
          Irene Cruz-Gonz\'alez \inst{6}
          }

   \institute{Instituto Nacional de Astrof\'isica, \'Optica y Electr\'onica, Luis Enrique Erro $\#1$, Tonantzintla, Puebla 72840, M\'exico
             \and
             Max-Planck-Institut f\"ur Radioastronomie, Auf dem H\"ugel 69, D-53121 Bonn, Germany
             \and
             I. Physikalisches Institut, Universit\"at zu K\"oln, Z\"ulpicher Strasse 77, K\"oln, Germany
             \and
             Byurakan Astrophysical Observatory after V.A. Ambartsumian, Aragatsotn Province 378433, Armenia
             \and
             Astrophysical Research Laboratory of Physics Institute, Yerevan State University, 1 Alek Manukyan St., Yerevan, Armenia
             \and
             Universidad Nacional Aut\'onoma de M\'exico, Instituto de Astronom\'ia, AP 70-264, CDMX 04510, Mexico
             }

   \date{Received May 7, 2025; accepted November 10, 2025}

  \abstract
   {}
   {This study investigates the relationship between the Mg II $\lambda$2798 \AA\ emission line and the 3000 \AA\ continuum luminosity, as well as the Baldwin Effect, in a sample of 40,685 radio-quiet (RQ) quasars and 441 Flat Spectrum Radio Quasars (FSRQs).}
   {We perform a comprehensive re-evaluation of the Mg II–3000 \AA\ correlation, explicitly accounting for dispersion introduced by AGN variability. After excluding >3000 radio-loud sources, we employ a binning technique to mitigate variability effects, yielding a refined empirical relation. We also further examine the Non-Thermal Dominance (NTD) parameter, to investigate the dominant source of the continuum.}
   {Our analysis reveals statistically significant differences in the slopes of the line–continuum luminosity relation between RQ quasars and FSRQs, with a parallel discrepancy in the Baldwin Effect. These findings imply either (1) intrinsic differences in the accretion disk spectra of RQ AGNs and FSRQs or (2) jet-induced continuum emission in FSRQs contributing to Broad Line Region (BLR) ionization. We also found that a substantial fraction of both RQ quasars (43.8\%) and blazars (55.5\%) exhibit NTD < 1. For blazars, this suggests that the accretion disk alone cannot fully explain BLR ionization; while we interpret $NTD < 1$ in radio-quiet quasars as a signature of several physical mechanisms: anomalies in the BLR structure (such as outflow or inflows), time lags between continuum and line variations, and the suppression of the UV continuum by a strong corona that diverts accretion power. Finally, we demonstrate that the Baldwin Effect emerges naturally from the line–continuum luminosity relationship, requiring no additional physical mechanism to explain its origin.
}
   {}


\titlerunning{Mg II and Continuum Luminosity in Blazars}

   \maketitle
%

\section{Introduction} \label{introduction}

A well-established correlation exists between the luminosity of broad emission lines and their corresponding continuum in active galactic nuclei (AGN). This is especially evident for line-continuum pairs such as H$\beta$-5100 \AA, Mg II $\lambda2798~$\AA-3000 \AA, and C IV $\lambda1549~$\AA-1350 \AA\ \citep[e.g.][among others]{Greene2005, Shen2011,Rakshit2020}. This relationship is expected, as the emission line luminosity increases with the ionizing continuum, assuming sufficient material in the broad-line region (BLR) is available for ionization.

\citet{Baldwin1977} was the first to identify a relationship between the rest-frame equivalent widths of ultraviolet emission lines (e.g., C IV, Ly$\alpha$) and the continuum luminosity at 1350 \AA\,(L$_{1350}$), a phenomenon now known as the \textit{Baldwin Effect} (hereafter BE), as named by \citep{Carswell1978}. The BE has since been the focus of numerous studies, as it provides valuable insights into the properties of various emission regions in active galactic nuclei. The BE has been well established for ultraviolet and optical broad emission lines \citep[e.g.,][]{Shields2007,Dong2009b,Bian2012,Sameshima2020,Xiao2022a}, as well as for narrow emission lines \citep[e.g.,][]{Croom2002, Dietrich2002, Netzer2004, Netzer2006, Netzer2007, Kovacevic2010, Popovic2011, Zhang2013}.

Over the past three decades, numerous studies have sought to identify the physical mechanisms underlying the BE. One widely accepted empirical explanation suggests that the ionizing continuum softens as luminosity increases \citep[e.g.,][]{Zheng1993}. Additionally, theoretical models propose that the BE is influenced, at least in part, by the shape of the continuum and the gas metallicity \citep[e.g.,][]{Korista1998}. Furthermore, several authors \citep[e.g.,][]{Zheng1993,Zheng1995,Espey1999,Dietrich2002} have found that emission lines from higher-ionization ions exhibit steeper slopes in BE diagrams. Other possible factors that may contribute to the BE include the Eddington ratio \citep{Baskin2004,Bachev2004,Dong2009} and black hole mass \citep[e.g.,][]{Warner2003,Xu2008}, though no consensus has been reached on these aspects.

\cite{PatinoAlvarez2016} proposed that the BE is a natural consequence of the relationship between emission line and continuum luminosities. By applying a change of variable, the logarithmic expression for the BE can be derived from the relationship between the luminosities. In other words, if a relationship exists between the luminosities of a continuum and a specific emission line, a BE will always arise for that emission line. If this hypothesis holds, the difference in the slopes of both relations (for the same emission line and continuum) should equal one \citep[see Section 4 of][]{PatinoAlvarez2016}. For the limited sample studied in that work, this condition is met for the line-continuum pairs previously mentioned.

This paper has several objectives: (1) Test the hypothesis proposed by \cite{PatinoAlvarez2016} using large samples of both radio-quiet (RQ) AGN and flat-spectrum radio quasars (FSRQ), where measurements of the Mg II $\lambda$2798 \AA\ emission line and the 3000 \AA\ continuum are possible. (2) Re-examine the relationship between this emission line and continuum for RQ quasars, as previously established by \cite{Shen2011}. (3) Investigate whether the Non-Thermal Dominance (NTD) parameter affects the slopes of the line-continuum relation or the BE. (4) Identify whether there are sources (either RQ or FSRQ) with NTD values below one, which may suggest that the accretion disk is not the sole provider of ionizing photons for the BLR.

The distinction between the two samples is of significant interest, as several studies over the past few decades have highlighted differences in the spectral properties of AGN based on radio-loudness. For example, a composite spectrum of a sample of radio-loud AGN shows stronger [O III] emission, broader Balmer lines, a redder spectral energy distribution (SED), and a stronger red wing weaker blue wing asymmetry in the C IV\,$\lambda$1549 \AA\ emission line, compared to an analogous spectrum for RQ AGN \citep[][and references therein]{Brotherton2001}.

The Radio-loudness classic criteria, where $R$ is the ratio between the radio (5\,GHz) and optical (4400\,\AA) flux densities, $R=F _{\mathrm{5\,GHz}}/F\,_{\mathrm{4400\,A}}$ \citep{Kellermann1989, Shastri1993}, separates radio-loud and radio-quiet sources at $R=10$. 

The structure of the paper is as follows: In Sect.~\ref{sec:sample}, we present the characteristics of the RQ AGN and FSRQ samples, along with the parameter calculations. Sect.~\ref{sec:relations} describes the fitting procedure for the luminosity-luminosity relations and the BE. In Sect.~\ref{sec:NTD}, we describe the calculation of the NTD parameter, its interpretation, and the results for the samples separated by NTD. In Sect.~\ref{sec:BE_origin}, we test the hypothesis that the BE is a natural consequence of the relationship between emission line luminosity and continuum luminosity. We discuss our results in Sect.~\ref{sec:Discussion}. Finally, the conclusions are summarized in Sect.~\ref{sec:conclusions}.

Throughout the paper a flat cosmology model is used with parameters $\Omega_{m}=0.3$, $\Omega_{\Lambda}=0.7$, and $H_0=70$ km\,s$^{-1}$\,Mpc$^{-1}$, in order to compare with \cite{Shen2011}.

\section{The samples}
\label{sec:sample}

In this study, we use two different samples: a control sample derived from the Sloan Digital Sky Survey (SDSS) quasar catalog \citep[][hereafter S11]{Shen2011} and an FSRQ-type blazar sample from the 5th Roma-BZCAT blazar catalog \citep{Massaro2015}.

\subsection{Control sample dominated by radio-quiet AGNs}
\label{rq-sample} 

To construct a control sample for comparison with our blazar-only sample, we extracted data for the Mg II $\lambda$2798 \AA\ emission line, the 3000 \AA\ continuum luminosities, and the Mg II equivalent width from the SDSS Quasar Catalog (S11). We applied the same filtering criteria as S11, selecting only sources with uncertainties in both luminosities lower than 0.03 dex. This resulted in an initial sample of approximately 44,000 sources. However, when plotting $\log(\lambda L_{3000})$ versus $\log(L_{Mg\, II})$, we identified a number of outlying sources that deviated from the general trend. To address these, we first inspected the spectra of a subset of clear outliers and identified issues with the quality of the spectra, mainly, the Mg II emission line was either absent, cropped, or misidentified, which explained the anomalous luminosity values. We then calculated the orthogonal distance of each source from the best-fit line to the unbinned sample ($\sim$44,000 points). Starting with the largest distances, we visually inspected their corresponding individual spectrum in the SDSS database. The 101 sources with the largest orthogonal distances all showed spectral problems. After excluding these outliers, we were left with a sample of 43,756 quasars.

We applied a linear fit to the relationship between the Mg II $\lambda$2798 \AA\ emission line and the 3000 \AA\ continuum luminosities for the resulting sample, following the same methodology as S11, using the BCES bisector and (Y|X) fitting techniques. Our fit yielded the same parameters as those presented in S11 (equations 11 and 12), confirming that our sample is consistent with the one used in S11. However, we noticed that the p-value of the fit was 1 (to machine precision), whereas the typical threshold for a reliable fit is 0.05. This indicates that the linear regression (and consequently the fit presented in S11) does not adequately represent the data.

The primary reason for the scatter, and consequently the low significance of the fit, is the inherent variability of AGNs. Since the timing of the SDSS observations does not account for the AGN's activity state, the observations could capture the AGN in a low-activity, high-activity, or intermediate state. This variability introduces additional scatter into the relationship, making the linear fit statistically unreliable. To address this limitation without acquiring additional data, we propose sorting the data by continuum luminosity and creating bins. The rationale behind this approach is that each bin, containing a sufficient number of data points, will include a balanced mix of objects in both low and high activity states. As a result, the variability should largely cancel out within each bin. We will apply this method in Section~\ref{sec:relations} to reevaluate the relationship between the Mg II $\lambda$2798 \AA\ emission line and the 3000 \AA\ continuum luminosities for RQ AGNs.

\subsection{FSRQ-type blazars}
\label{fsrq-sample} 

The fifth edition of the Roma-BZCAT catalog (Massaro et al. 2015) is one of the most comprehensive for blazar identification.

To construct a sample of FSRQ-type blazars, we used the 5th Roma-BZCAT (5BZCAT) catalog of blazars \citep{Massaro2015}. This catalog is one of the most comprehensive for blazar identification and comprises both confirmed blazars and sources displaying characteristics consistent with this type, ensuring a reliable selection for our study. From this catalog, we extracted all sources designated as FSRQ and then searched for available spectra of these objects in the Sloan Digital Sky Survey (SDSS) Data Release 16 \citep[DR16,][]{Ahumada2020}. Upon reviewing the spectra, we found that the maximum wavelength coverage ranged from 3650 to 10400 \AA. Since we are focused on the Mg II $\lambda$2798 \AA\ emission line and the corresponding 3000 \AA\ continuum, we limited our sample to objects with redshifts between 0.34 and 2.41. After applying these filters, we were left with a final sample of 441 FSRQ-type blazars with the Mg II$\lambda$2798 \AA\ emission line in their spectra.

Although the 5BZCAT is a blazar catalog, we wanted to ensure that every source in our FSRQ sample is indeed a blazar. To verify this, we analyzed radio images of all the sources using data from the FIRST \citep{Becker1995} and LOFAR \citep{Shimwell2022} surveys, specifically looking for extended radio structures on kiloparsec scales to check for potential contamination by non-blazar sources. Additionally, we examined VLBI images from the Astrogeo VLBI FITS image database\footnote{\url{http://astrogeo.org/vlbi_images/}} to assess whether the jet is aligned on both parsec and kiloparsec scales. For sources with extended radio structures on kiloparsec scales, any misalignment would suggest the possibility of the pc-scale jet being blazar-like. We also considered optical light curves as an additional diagnostic tool, given that optical synchrotron emission is produced in the jet at the parsec scale. We retrieved light curve data from the Zwicky Transient Facility (ZTF) Database \citep{Bellm2019, Masci2019}. 

Using the aforementioned data, we classify sources as confirmed blazars if they exhibit flaring behavior in their optical light curves. These flaring events are characterized by high amplitude and short timescales, which are partially attributed to Doppler boosting, requiring a small jet viewing angle. However, it is possible that a source did not undergo a flaring event during the observation period. Therefore, we classify such sources as possible blazars if they exhibit a compact or one-sided morphology in radio images. Finally, sources that display a two-sided radio morphology, characteristic of radio galaxies, and do not show flaring behavior are classified as non-blazars. Further details of this analysis can be found in Appendix~\ref{non-blazars}.

\subsubsection{Spectra manipulation and fitting}

The spectra were transformed into the rest-frame, and the monochromatic flux was corrected by a factor of (1+z) to facilitate comparison with the RQ sample described in Section~\ref{rq-sample}. The wavelength range was then trimmed to 2500-3100 \AA\ to simplify the fitting process. We corrected the spectra for Galactic extinction using the dust maps from \cite{Schlafly2011}, which recalibrate the data from \cite{Schlegel1998}. The reddening law of \cite{Cardelli1989} with $R_V=3.1$ \citep{Fitzpatrick1999} was adopted for the extinction correction.

\begin{figure}[htbp]
\begin{center}
\includegraphics[width=0.5\textwidth]{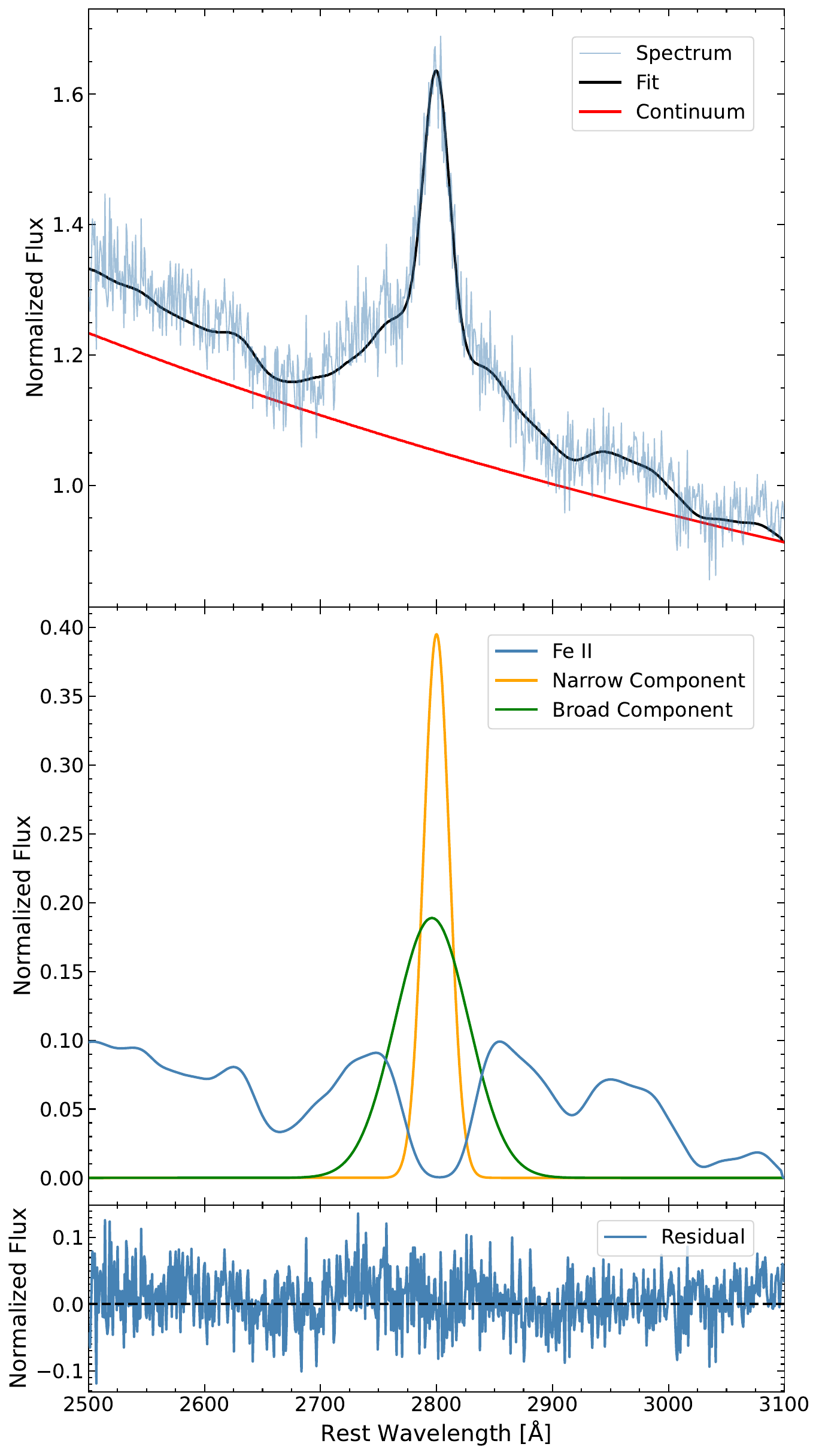}
\caption{Example of the fitting procedure for the spectra of the FSRQ sample. The source is 5BZQ J1520+0732 (SDSS J152045.54+073230.5). Top panel: The observed spectra (blue), the final fit (black), and the power-law continuum (red). Middle panel: Spectral components fitted (excluding the continuum), including the broad Gaussian component (green), the narrow Gaussian component (orange), and the fitted Fe II emission (blue). Bottom panel: Residuals from the fit.}
\label{fitting}
\end{center}
\end{figure}

We then performed spectral decomposition using the IRAF task \texttt{specfit}, where the continuum, Fe II emission, and emission lines were modeled simultaneously. The spectral continuum was modeled as a power-law. Initial parameter guesses for the continuum (normalization and slope) were selected interactively by visually inspecting the spectrum. The Fe II emission was fitted using the template from \cite{Vestergaard2001}, with adjustments to the intensity and broadening to match each individual spectrum. The Mg II $\lambda 2798$ \AA\ emission line was modeled using two Gaussian components, although in some cases, a single Gaussian was sufficient to capture the entire line profile. Since the spectra lacked a narrow-line region, we were unable to fix the FWHM for the narrow-line component of the Mg II emission line. As a result, we focused on carefully fitting the line profile. The final parameters, including the amplitude of the power-law continuum component, were then determined through a non-linear least-squares minimization (Levenberg-Marquardt algorithm) of the combined model against the observed data. This simultaneous fitting method accounts for the inherent covariance between the continuum and emission line fluxes. An example of the fitting procedure for two different sources is shown in Fig.~\ref{fitting}.

\subsubsection{Flux and EW measurements}\label{flux_measure}

The spectral continuum flux at 3000 \AA\ was measured as the mean value within the wavelength range 2950–3050 \AA, from the spectra after subtracting the Fe II emission and the Gaussian components representing the Mg II emission line. We considered two sources of uncertainty: the first is the signal-to-noise ratio of the spectrum, which we quantified as the standard deviation of the observed spectrum (after subtracting the Fe II and Gaussian components) within the same wavelength range. The second source of uncertainty is the calibration error, which was derived by averaging the error spectrum from SDSS (inverse variance) in the same wavelength range. We combined these two contributions in quadrature to obtain the total uncertainty in the 3000 \AA\ continuum flux.

\begin{figure}[htbp]
\begin{center}
\includegraphics[width=0.45\textwidth]{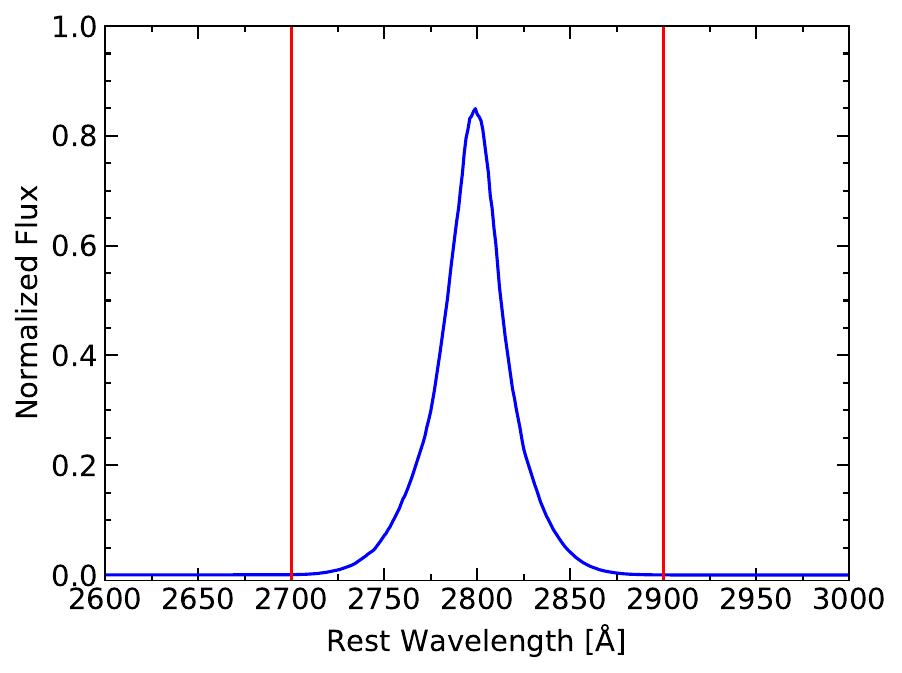}
\includegraphics[width=0.45\textwidth]{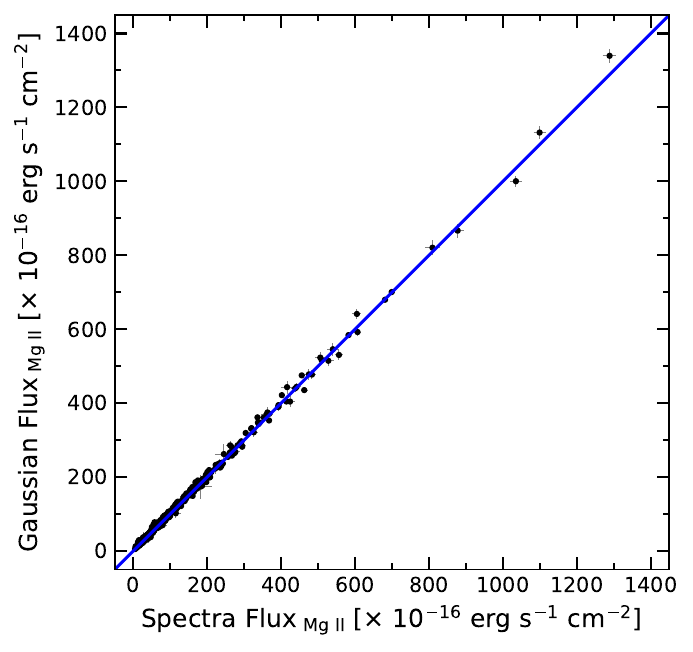}
\caption{Top: The median of the emission line components (blue), with red vertical lines indicating the integration limits. Bottom: Comparison of the emission line fluxes obtained by integrating the spectra (after subtracting the continuum and Fe II emission) versus integrating the fitted Gaussians. The blue line represents the $y = x$ equality.}
\label{emissionline}
\end{center}
\end{figure}

The flux of the Mg II$\lambda 2798$ \AA\ emission line was measured by integrating the spectrum after subtracting the continuum and Fe II emission. To define the wavelength range for integration, we created a composite median profile (see top panel of Fig.~\ref{emissionline}) and determined that the integrated emission line flux should be measured in the 2700–2900 \AA\ range. We also tested whether there was a significant difference between directly integrating the spectra or integrating the Gaussian components (see bottom panel of Fig.~\ref{emissionline}) and found no significant difference.

For the uncertainty in the integrated flux of the Mg II emission line, we considered three different sources. The first source of uncertainty comes from the Fe II emission, as it is blended with the Mg II emission line. When subtracting the Fe II component, the line flux can be affected. This contribution was calculated following \cite{LeonTavares2013} using Eq.~\ref{errorfe}. In this equation, $\sigma$ represents the root mean square (rms) of the continuum after the Fe II subtraction (in a region where there is no emission line but Fe II emission is present), $max_s$ is the maximum value of the emission line after subtracting both the Fe II and the continuum, $F_{line}$ is the integrated emission line flux, and $F_{ironless}$ is the fraction of the emission line flux within the integration range where no iron subtraction occurs (i.e., where the Fe II template is zero).

\begin{equation} \label{errorfe}
\sigma_{Fe}=(\sigma/max_s)(F_{line}-F_{ironless})
\end{equation}

The second source of uncertainty is the signal-to-noise ratio (S/N) of each individual spectrum. We calculate this contribution following the method described in \cite{Tresse1999}, using Eq.~\ref{errortresse}, which accounts for the noise in the spectrum and its impact on the measurement of the emission line flux.

\begin{equation} \label{errortresse}
\sigma_F=\sigma_c D \sqrt{ 2N_{pix} + EW/D }
\end{equation}

Here, $\sigma_c$ denotes the mean standard deviation per pixel of the continuum on either side of the emission line, $D$ represents the spectral dispersion in \AA\ per pixel, $N_{pix}$ is the number of pixels spanned by the emission line, and EW is the equivalent width of the emission line. 

The third source of uncertainty is the flux calibration error, obtained from the SDSS error spectra. To calculate the total uncertainty in the emission line flux, we combine the three sources of uncertainty — Fe II subtraction, signal-to-noise ratio, and flux calibration error — by adding them in quadrature

The luminosities of the Mg II $\lambda 2798$ \AA\ emission line (${\rm L_{MgII}}$) and the 3000 \AA\ spectral continuum (${\rm \lambda L_{\lambda 3000}}$) are calculated using the luminosity distance for each individual source, following the same cosmology as in S11 (referenced at the end of Section~\ref{introduction}) to ensure a direct comparison of luminosity values. The uncertainties in the luminosities were determined through error propagation.

The equivalent width (EW) of the Mg II emission line is calculated by integrating the spectrum alongside the fitted continuum emission, as shown in Eq.~\ref{ewmeasure}. Here, $F_{\lambda}$ represents the monochromatic flux of the spectrum after subtracting the Fe II emission, $F_c$ is the fitted continuum flux, and $\lambda_1$ and $\lambda_2$ are the boundaries of the integration range (2700-2900 \AA).

\begin{equation} \label{ewmeasure}
EW = \int_{\lambda_1}^{\lambda_2}{\frac{ F_{\lambda} - F_c }{ F_c } } d\lambda
\end{equation}

The uncertainty in the equivalent width is determined using Eq.~\ref{errorew} from \cite{Tresse1999}, where $F$ represents the emission line flux. The remaining parameters are consistent with those defined in Eq.~\ref{errortresse}.

\begin{equation} \label{errorew}
\sigma_{EW}=\frac{EW}{F}\sigma_c D \sqrt{ EW/D + 2N_{pix} + (EW/D)^2/N_{pix} }
\end{equation}

\begin{figure*}[htbp]
\begin{center}
\includegraphics[width=0.95\textwidth]{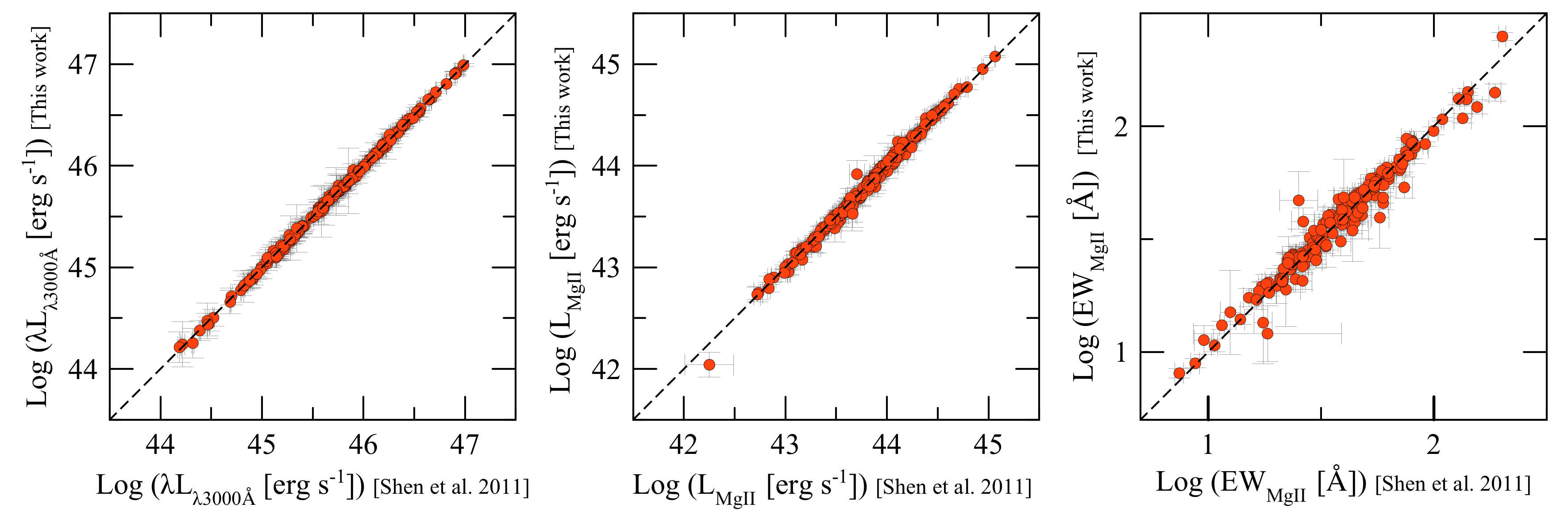}
\caption{Comparison between the measurements (for the common sources) performed in this work and in S11, for the 3000 \AA\ continuum luminosity (left panel), the Mg II $\lambda 2798$ \AA\ emission line luminosity (middle panel), and the Mg II $\lambda 2798$ \AA\ equivalent width (right panel). The dashed line represents $y = x$.}
\label{comparison}
\end{center}
\end{figure*}

Finally, to assess whether our measurement procedure aligns with that of S11, we cross-referenced our FSRQ sample with the SDSS Quasar Catalogue, identifying 183 matching sources. The average differences for emission line luminosity, continuum luminosity, and equivalent width were found to be 0.014, 0.036, and 0.043 dex, respectively. Since these differences are comparable to or smaller than the typical uncertainties in these quantities, we conclude that both measurement methods are consistent. A comparison of these parameters is shown in Fig.~\ref{comparison}. Additionally, we observed that the uncertainties estimated in this work are generally larger than those reported in the SDSS Quasar Catalogue by S11, suggesting that the uncertainties in the latter may be underestimated.

\section{Line and continuum luminosity relations and the Baldwin effect} \label{sec:relations}

\subsection{Radio-quiet control sample}

\begin{figure*}[htbp]
\begin{center}
\includegraphics[width=0.95\textwidth]{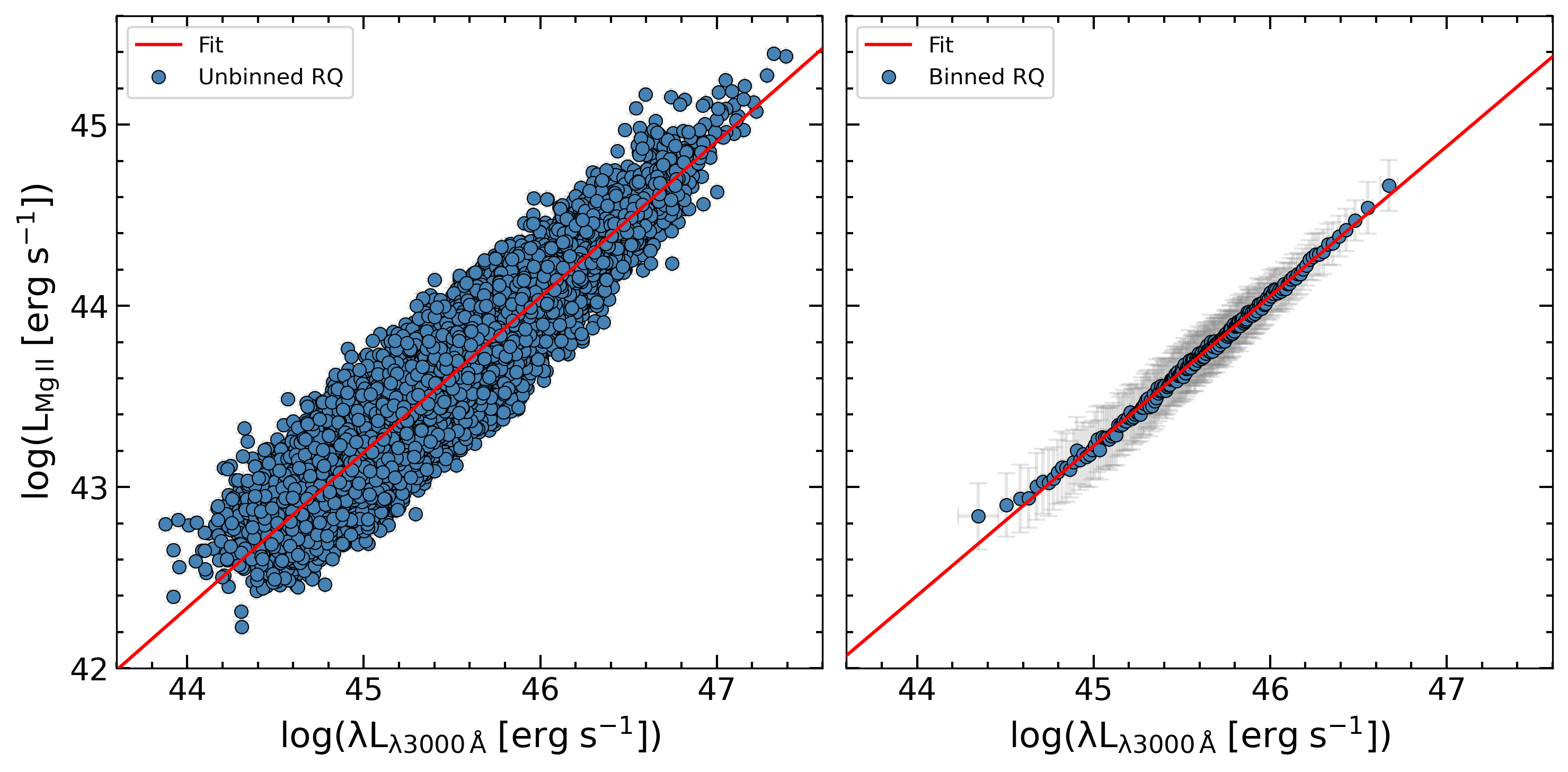}
\caption{The relationship between the 3000 \AA\ continuum luminosity and the Mg II $\lambda$2798 \AA\ emission line luminosity for the RQ control sample is shown, with the unbinned data in the left panel and the binned data in the right panel. The red solid line represents the fitted relation for each data set. In the right panel, the fitted line corresponds to the relationship presented in Eq.~\ref{eqlumirelRQ}.}
\label{lumirelRQ}
\end{center}
\end{figure*}

To ensure that our control sample was not strongly influenced by jet contributions to the continuum emission, we filtered out radio-loud sources (radio-loudness $>10$) using the radio-loudness values from the S11 quasar catalog. This resulted in a final sample of 40,685 RQ sources. We then determined the number of points per bin as $\sqrt{n}$, resulting in 202 bins, each containing 201 points. For each bin, we computed the weighted mean of the logarithmic values for the Mg II emission line luminosity, the 3000 \AA\ continuum luminosity, and the Mg II equivalent width (the latter of which will be used later). We observed that the standard deviation within each bin for all three parameters was significantly higher than the standard error of the weighted mean. To remain conservative, we used the standard deviations as error bars for the binned data. To perform the linear fit, we used the Python procedure ODR \citep{Boggs1990}, a maximum-likelihood estimator of model parameters that accounts for errors in both axes. The resulting fit for the Mg II - continuum luminosity relationship produced a p-value of zero (to machine precision), indicating an accurate model representation, with a Pearson correlation coefficient of 0.998 (p-value of $10^{-235}$ for the correlation). The fit is presented in Eq.~\ref{eqlumirelRQ} and shown in the right panel of Fig.~\ref{lumirelRQ}. We also attempted a linear fit using the unbinned data after filtering out the radio-loud sources; however, the p-value of this fit was one (to machine precision), similar to the result in S11. This fit is shown in the left panel of Fig.~\ref{lumirelRQ}. It is important to note that the first and last bins, representing the lowest and highest luminosities, may predominantly contain objects in low and high activity states, respectively. This likely accounts for the slight separation of these points (still within the error bars) from the overall fit.

\begin{equation} \label{eqlumirelRQ}
\log{L_{MgII}} = (0.826\pm0.025)\log{\lambda L_{\lambda 3000}} + (6.057\pm0.164)
\end{equation}

\begin{figure*}[htbp]
\begin{center}
\includegraphics[width=0.95\textwidth]{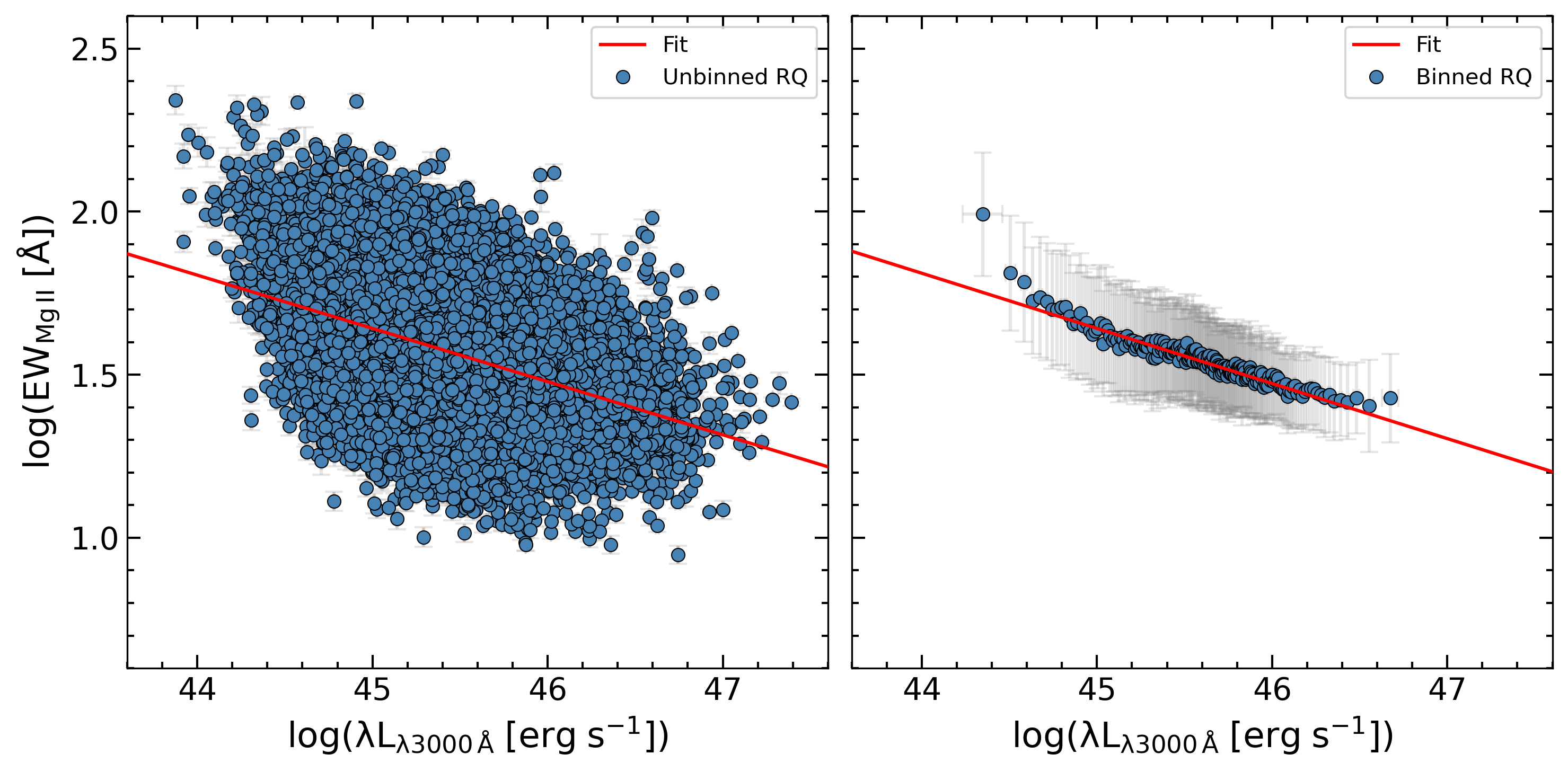}
\caption{Relationship between the 3000 \AA\ continuum luminosity and the equivalent width of the Mg II $\lambda$2798 \AA\ emission line for the RQ control sample, with the unbinned data shown in the left panel and the binned data in the right panel. The solid red line in both panels represents the fitted relation obtained for each dataset.}
\label{BERQ}
\end{center}
\end{figure*}

Using the same RQ sample, binning method, and fitting routine as described previously, we tested for the BE relationship. The results are presented in Figure~\ref{BERQ}. The resulting linear fit, shown in Eq.~\ref{eqBERQ}, has a p-value of zero (to machine precision) and a Pearson correlation coefficient of -0.948, with a p-value of $10^{-101}$ for this correlation coefficient.

\begin{equation} \label{eqBERQ}
\log{EW_{MgII}} = (-0.169\pm0.024)\log{\lambda L_{\lambda 3000}} + (9.247\pm0.172)
\end{equation}

\subsection{FSRQ sample}

With all luminosities and equivalent widths calculated and confirmed to be consistent with the S11 measurements, we can now explore the relationship between the continuum and emission line luminosities, as well as the BE for our sample of FSRQs. Similar to the RQ control sample, an initial error-weighted linear least-squares fit (considering errors on both axes) for all 441 sources yielded a fit with a p-value of 1 (to machine precision). This result is expected, as FSRQs are known to exhibit even greater variability than RQ AGN \citep[e.g.][]{Ulrich1997,Beckmann2007}. Consequently, we applied the same binning methodology used for the RQ control sample, which produced 21 bins containing 21 data points each.

\begin{figure*}[htbp]
\begin{center}
\includegraphics[width=0.95\textwidth]{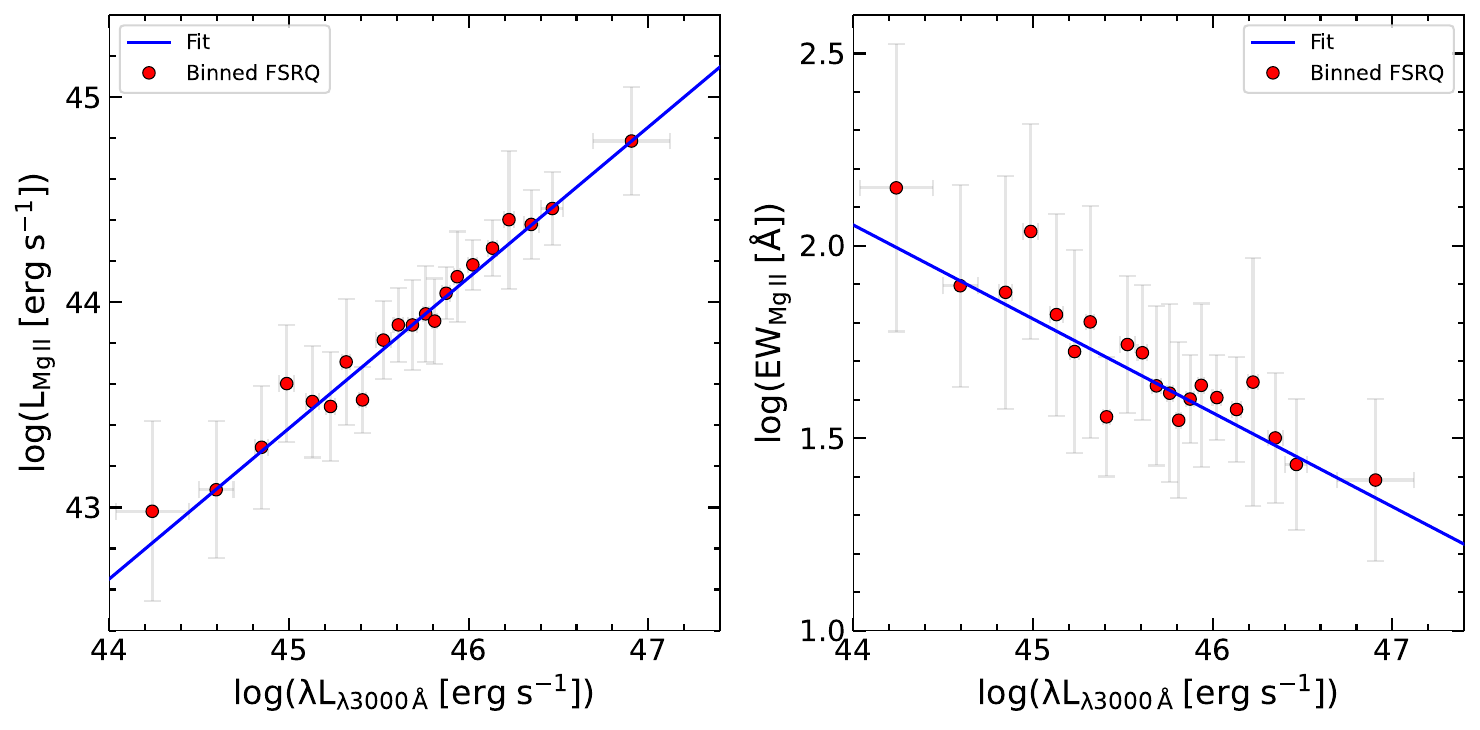}
\caption{Left panel: Relationship between the 3000 \AA\ continuum luminosity and the Mg II $\lambda 2798$ \AA\ emission line luminosity for the binned data of the FSRQ sample. Right panel: BE relationship for the binned data of the FSRQ sample. In both panels, the solid line represents the fit to the binned data.}
\label{lumirelFSRQ}
\end{center}
\end{figure*}

The fit for the relationship between the Mg II $\lambda 2798$ \AA\ emission line luminosity and the 3000 \AA\ spectral continuum luminosity yields a p-value of $2.3 \times 10^{-6}$, with a Pearson correlation coefficient of 0.986 (with a corresponding p-value of $3.8 \times 10^{-16}$ for this correlation). This relationship is presented in Eq.~\ref{eqlumirelFSRQ} and is illustrated in the left panel of Fig.~\ref{lumirelFSRQ}.

\begin{equation} \label{eqlumirelFSRQ}
\log{L_{MgII}} = (0.734\pm0.095)\log{\lambda L_{\lambda 3000}} + (10.356\pm1.571)
\end{equation}

Similarly, for the BE the linear fit yields a p-value of $3.8 \times 10^{-6}$, with a Pearson correlation coefficient of -0.913 (and a corresponding p-value of $7 \times 10^{-9}$ for this correlation). This relationship is presented in Eq.~\ref{eqBEFSRQ} and is shown in the right panel of Fig.~\ref{lumirelFSRQ}.

\begin{equation} \label{eqBEFSRQ}
\log{EW_{MgII}} = (-0.244\pm0.083)\log{\lambda L_{\lambda 3000}} + (12.790\pm1.424)
\end{equation}

\subsection{Slope comparisons} \label{sec:slopes}

\cite{PatinoAlvarez2016} conducted simulations showing that variability, whether from the accretion disk or the jet, should not alter the slope of the continuum luminosity–line luminosity relationship, as long as the accretion disk remains the sole source of ionization for the BLR. Based on this, we compared the slopes obtained for this relation in both the RQ and FSRQ samples. We applied a Welch unpaired t-test to the slope values and their uncertainties, using the appropriate degrees of freedom (calculated with the Welch formula due to unequal variances). The results yielded a p-value of $2.6 \times 10^{-4}$ for the continuum luminosity–line luminosity relationships and a p-value of $5.3 \times 10^{-4}$ for the BE. This indicates a statistically significant difference in the slopes of these relationships between RQ AGN and FSRQ-type blazars.

Given that the aforementioned simulations suggest the difference in slope cannot be explained solely by variability from either the accretion disk or the jet, three possibilities remain: (1) The ionizing spectrum of accretion disks in FSRQ-type blazars may be significantly and systematically different from that in RQ AGN, a hypothesis that has not been extensively explored in the literature; (2) In blazars, part of the broad emission line photons are used to produce gamma-rays via inverse Compton scattering; (3) The jet contributes to the ionization of the broad-line emitting material.

\cite{Xiao2022a} mention that the correlation obtained between the equivalent width of the Mg II line, and the intrinsic inverse Compton luminosity is evidence of scenario number (2), however, the correlation coefficient of 0.27, with a coefficient of determination ($r^2$) of 0.073, implies that only 7.3\% of the variability in the EW$_{Mg~II}$ is related to the variability in the inverse Compton luminosity. To explore whether these values of coefficient of determination are significant, we took the 4FGL DR4 point source catalog (version 34), filtered the FSRQ sources, and calculated the error percentage in the energy flux, for all the FSRQ sources in the catalog, we found that the average uncertainty is 11.3\%, which is higher than the coefficients of determination mentioned above, therefore, we don’t think this can be considered a true correlation. Therefore, we cannot support this correlation as evidence of the scenario. However, we do not discard the possibility that the loss of BLR photons to inverse Compton scattering is a contributing factor in the difference between the slopes of radio-quiet AGN and FSRQ.

On the other hand, prior observational evidence supports the occurrence of scenario (3) in blazars \citep[e.g.][]{LeonTavares2013, Isler2013, Chavushyan2020, AmayaAlmazan2021, Hallum2022, AmayaAlmazan2022}. Therefore, we propose that the observed difference in the slope of the continuum luminosity–line luminosity relation between RQ AGN and FSRQ-type blazars is due, at least in part, to jet-driven ionization of the broad-line emitting material.

\section{Non-thermal dominance} \label{sec:NTD}

The Non-Thermal Dominance (NTD) parameter \citep{Shaw2012} is a diagnostic tool used to quantify the contribution of jet synchrotron emission to the optical/UV continuum relative to the accretion disk. The NTD is defined in Eq.~\ref{NTDeq1}:

\begin{equation} \label{NTDeq1}
NTD=\frac{L_{obs}}{L_p}
\end{equation}

Here, $L_{obs}$ is the directly observed continuum luminosity, and $L_p$ is the predicted luminosity of the purely thermal accretion disk component. The predicted luminosity $L_p$ is derived by applying an empirical relationship—calibrated on a radio-quiet (RQ) AGN sample—that connects the luminosity of an emission line (Mg II in this case) to the luminosity of the ionizing continuum at 3000 Å. This RQ calibration sample, by definition, has negligible jet contribution, meaning the relation $L_{line} = f(L_{3000})$ defines the behavior of a canonical disk-ionized broad-line region.

Therefore, for a target source, $L_p$ represents the expected continuum luminosity if the BLR were solely illuminated by an accretion disk identical to those in RQ AGNs. Comparing this prediction ($L_p$) to the actual observation ($L_{obs}$) isolates the non-thermal jet contribution (see Eq.~\ref{NTDeq2}). The uncertainty in the NTD is calculated via standard error propagation.

\begin{equation} \label{NTDeq2}
NTD=\frac{L_{disk}+L_{jet}}{L_p}=\frac{L_{disk}+L_{jet}}{L_{disk}}=1+\frac{L_{jet}}{L_{disk}}
\end{equation}

Following this logic, the NTD can be interpreted as follows: a) For $1<NTD<2$, the observed continuum is consistent with being mostly thermal (disk-dominated). b) For $NTD > 2$, the observed continuum is more than double the thermal prediction, indicating a contribution from the jet synchrotron emission larger than that of the accretion disk. c) For $NTD < 1$, the observed continuum is fainter than predicted from the line luminosity; the physical interpretation of this result differs for Flat Spectrum Radio Quasars and radio-quiet AGNs, as discussed below.

In Section~\ref{rq-sample}, we revisited the S11 relation between the Mg II $\lambda 2798$ \AA\ emission line luminosity and the 3000 \AA\ continuum luminosity and recalculated it after filtering out the radio-loud objects from the same sample and applying binning to the data. While we could solve Eq.~\ref{eqlumirelRQ} for $\log{\lambda L_{\lambda 3000}}$ and propagate the associated errors to obtain the continuum luminosity, this approach could lead to an overestimation of the uncertainties in the relation's coefficients. To mitigate this, we refitted the data using the emission line luminosity as the independent variable. The resulting fit is presented in Eq.~\ref{newlumrel}. Upon inspection, the coefficients are identical to those obtained by solving Eq.~\ref{eqlumirelRQ} for $\log{\lambda L_{\lambda 3000}}$, but this method prevents overestimation of their uncertainties. The fit results in a p-value of zero (to machine precision). Eq.~\ref{newlumrel} is used to calculate the NTD parameter throughout the remainder of this paper.

\begin{equation} \label{newlumrel}
\log{\lambda L_{\lambda 3000} } = (1.211\pm0.036)\log{ L_{MgII} } + (-7.334\pm0.231)
\end{equation}

\begin{figure*}[htbp]
\begin{center}
\includegraphics[width=0.9\textwidth]{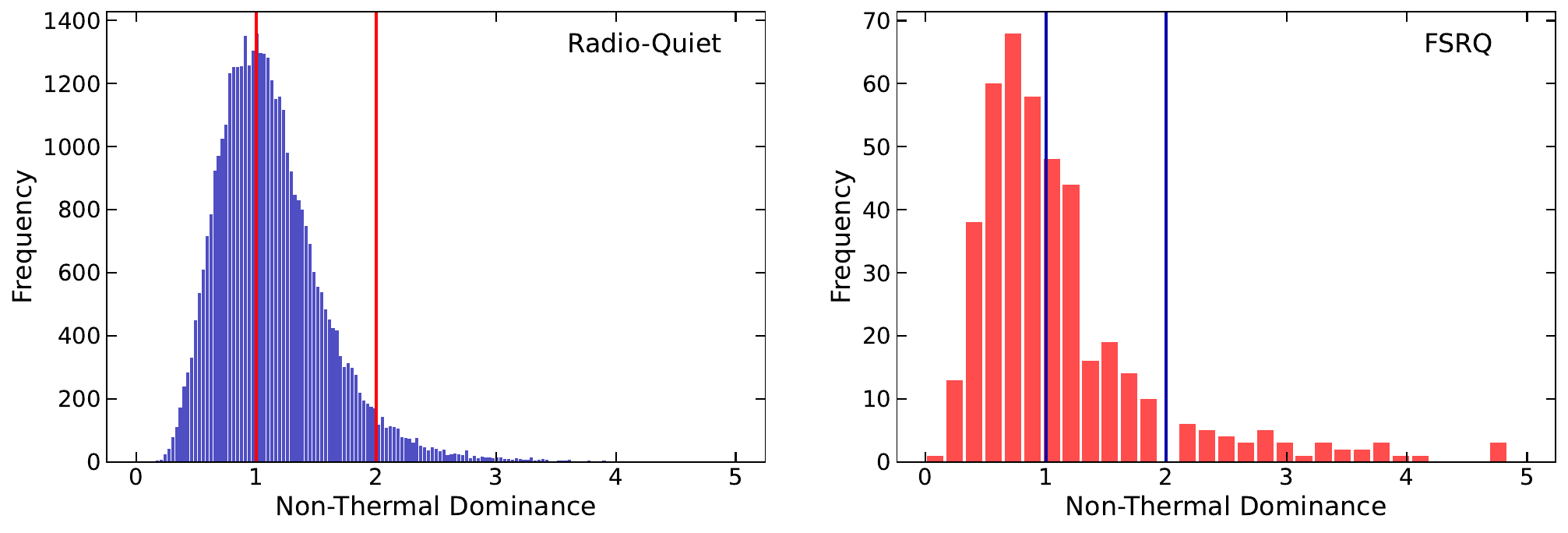}
\caption{Distributions of the NTD values for the RQ sample (left panel) and the FSRQ sample (right panel).}
\label{NTDdist}
\end{center}
\end{figure*}

We calculated the NTD values for both the control RQ sample and the FSRQ sample. The distributions are shown in Fig.~\ref{NTDdist}. The results reveal a complex picture. In the RQ sample, 21,047 sources (51.7\%) are disk-dominated ($1<NTD<2$), 1,818 (4.5\%) show signs of jet contribution ($NTD > 2$), and a significant fraction, 17,820 sources (43.8\%), exhibit an $NTD < 1$. In the FSRQ sample, 144 sources (32.7\%) are in a disk-dominated state ($1<NTD<2$), 52 (11.8\%) are jet-dominated ($NTD > 2$), and the majority, 245 sources (55.5\%), have $NTD < 1$.

The pronounced disparity in sample sizes ($\sim$2 orders of magnitude) between the datasets complicates a direct statistical comparison of their NTD distributions, as it can overpower goodness-of-fit tests like Kolmogorov-Smirnov (K-S) and Anderson-Darling (A-D), while undermining the rank-based Mann-Whitney U (MWU) test \citep{Hollander2014}. To facilitate a valid analysis, we compared the smaller sample against a representative subsample of the larger dataset, created via bootstrap resampling to preserve its statistical properties. The full analytical procedure, as well as the numerical results are detailed in Appendix~\ref{NTD_comparison}.

All three statistical tests consistently indicated that the NTD distributions of the full FSRQ sample and the RQ subsample are statistically significantly different, suggesting they originate from different parent populations. To investigate further, we stratified the samples into the different NTD regimes and repeated the bootstrap procedure and statistical testing. For the disk-dominated regime, all tests failed to reject the null hypothesis, indicating a common parent distribution for FSRQs and RQs. In contrast, for the jet-dominated regime, the tests revealed statistically significant differences. A significant difference was also found for sources with $NTD<1$.

The high number of sources with $NTD < 1$ is a robust result, persisting even when considering uncertainties, 38.6\% of RQ and 47.6\% of FSRQ sources have $NTD+\sigma_{NTD}<1$. This is not an artifact of the method but a meaningful diagnostic that points to several physical scenarios which we now explore.

\subsection{Interpretation of $NTD < 1$}
\label{NTD_lower_1}

As mentioned before, the interpretation of $NTD<1$ must be different whether we are talking about FSRQ or RQ AGN.

For FSRQs, the most plausible explanation for $NTD < 1$ is that the jet itself contributes to ionizing the BLR. The observed line luminosity ($L_{MgII}$) is high because it is powered by both the accretion disk and the jet's non-thermal radiation, leading to an over-prediction of the pure-disk luminosity ($L_p$). Evidence for this includes: a) The proposed existence of a jet-ionized line emitting region \citep{Arshakian2010b}. b) Direct observational evidence in individual blazars (e.g., 3C 454.3: \cite{LeonTavares2013, Isler2013, AmayaAlmazan2021}; CTA 102: \cite{Chavushyan2020}; B2 1633+382: \cite{AmayaAlmazan2022}; Ton 599: \cite{Hallum2022}). c) Scenarios where the base of the jet is embedded within the BLR \citep{LeonTavares2015, AmadorPortes2025}.

For RQ AGNs, the interpretation of $NTD < 1$ must differ, as significant jet ionization is unlikely. Multiple mechanisms intrinsic to the disk-BLR-corona system can explain a suppressed continuum relative to the line emission:

\begin{itemize}

\item Anomalies in the BLR. The BLR may not be a simple, virialized system. Geometrical or dynamical changes, such as strong outflows or inflows \citep[e.g.][]{Popovic2019}, can alter its reprocessing efficiency, leading to a different line-to-continuum ratio than the canonical RQ relation predicts.

\item Variability and Time Lags. The BLR responds to changes in the ionizing continuum with a time delay. A single-epoch observation could catch a fading continuum while the lines are still bright from a previous brighter state, resulting in a temporarily low NTD.

\item  Role of the Hot Corona. It is possible that the $NTD < 1$ reflects the energy budget of the disk-corona system, since a significant fraction of accretion power is dissipated in the hot corona via Comptonization of disk photons \citep[e.g.][]{Haardt1991, Risaliti2019}, shifting energy from the UV/EUV to the X-ray band. As a result, the observed UV continuum ($L_{obs}$) is suppressed relative to the total ionizing flux powering the broad-line region. Since the line luminosity ($L_{MgII}$) traces this total flux, it can exceed expectations based on $L_{ons}$ alone, producing an overestimated $L_p$ and thus $NTD<1$. This predicts an anti-correlation between corona strength (e.g., $\alpha_{OX}$) and NTD, which we will test in future work.

\end{itemize}

It is also important to note that a RQ classification (radio-loudness parameter $< 10$) does not absolutely preclude some level of jet activity. The presence of a small fraction (4.5\%) of RQ sources with $NTD > 2$ could be attributed to: a) Undetected radio flux leading to misclassification; b) A synchrotron peak at higher frequencies, enhancing optical non-thermal flux while reducing it at 5 GHz; c) The source being in a low state during the radio-loudness measurement; or d) The non-simultaneity of the optical and radio observations used to calculate the radio-loudness parameter \citep[e.g.][]{Kukula1998, Falcke2001, Liao2022}.

\subsection{Sub-sample analysis according to activity state}

Since the NTD values clearly distinguish the activity state of the sources, we decided to analyze subsamples of both the RQ and FSRQ samples, categorized by disk-dominated sources, jet-dominated sources, and objects with $NTD<1$. These subsamples were separated while considering the uncertainty in the NTD parameter. Disk-dominated sources are defined as those with $NTD - \sigma_{NTD} \geq 1$ and $NTD + \sigma_{NTD} < 2$. Jet-dominated sources are those satisfying $NTD - \sigma_{NTD} \geq 2$. Sources with $NTD+\sigma_{NTD}<1$ form their own separate subsample.

\begin{figure*}[htbp]
\begin{center}
\includegraphics[width=0.95\textwidth]{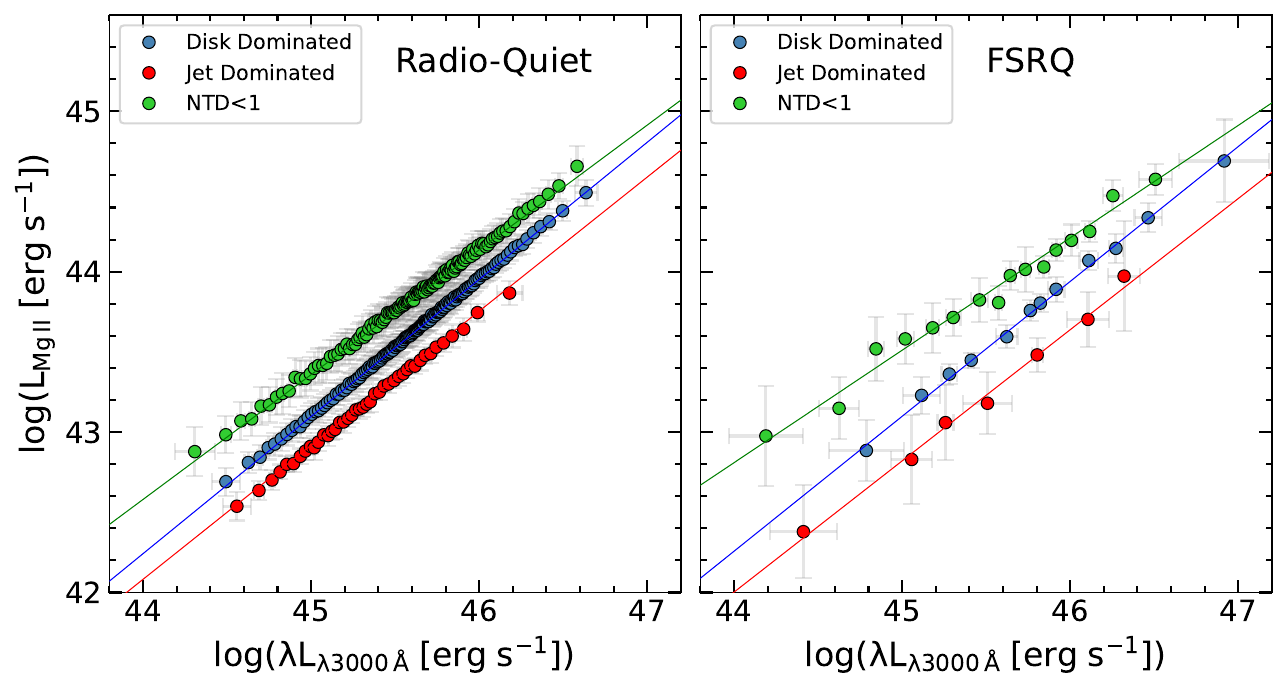}
\caption{Relationship between the 3000 \AA\ continuum luminosity and the Mg II $\lambda$2798 \AA\ emission line luminosity, using binned data, for disk-dominated (DD) sources (blue), jet-dominated (JD) sources (red), and objects with $NTD<1$ (green). The left panel presents the relations for the RQ sources, while the right panel shows the relations for the FSRQ sources.}
\label{Lum-Lum_NTD}
\end{center}
\end{figure*}

\begin{figure*}[htbp]
\begin{center}
\includegraphics[width=0.95\textwidth]{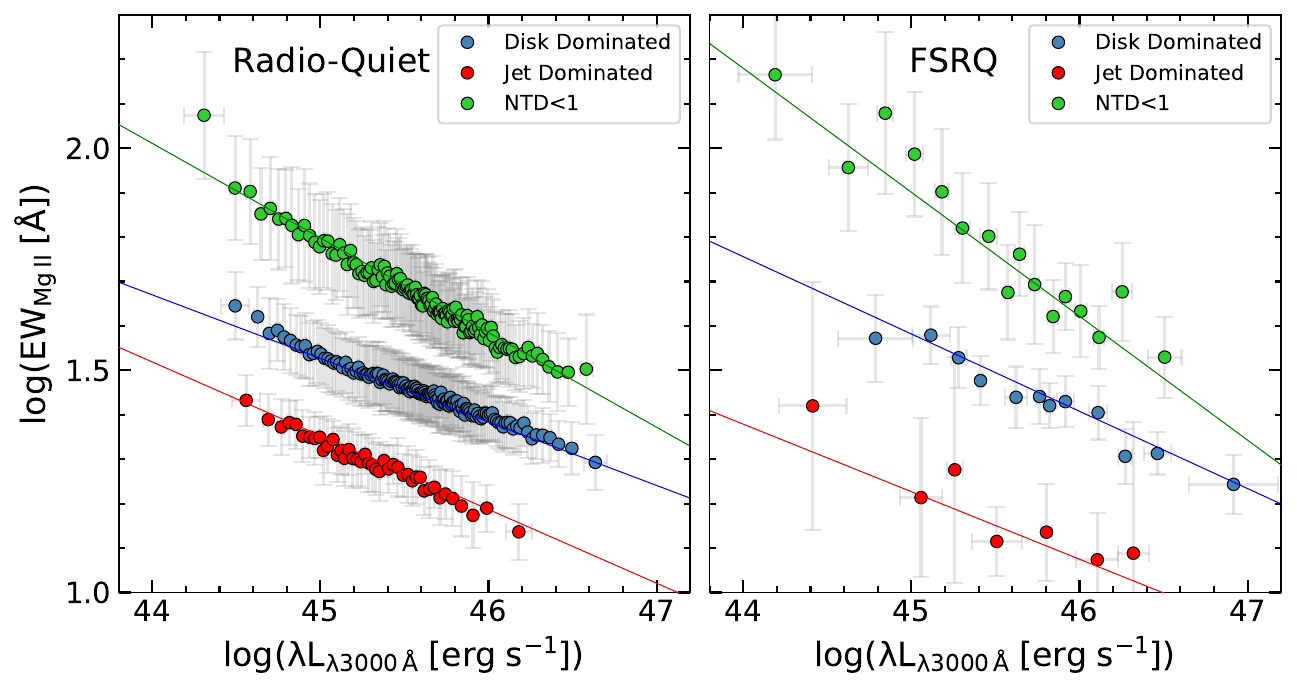}
\caption{BE for the 3000 \AA\ continuum and the Mg II $\lambda$2798 \AA\ emission line, using binned data, for disk-dominated (DD) sources (blue), jet-dominated (JD) sources (red), and objects with $NTD<1$ (green). The left panel displays the BE for the RQ sources, while the right panel shows it for the FSRQ sources.}
\label{BE_NTD}
\end{center}
\end{figure*}

As we did for the full samples, we attempted a linear fit of the data as is, but all the fits resulted in p-values approximately equal to 1. Therefore, we applied the same binning methodology to derive the continuum luminosity-line luminosity relation, as well as the BE, for these subsamples in both the RQ and FSRQ samples. The results are summarized in Table~\ref{results_NTD} and plotted in Figures~\ref{Lum-Lum_NTD} and \ref{BE_NTD}. The Pearson correlation coefficients and p-values for these relationships are provided in Table~\ref{correlations_NTD}.

\begin{table*}
\centering
\caption{Parameters obtained from the linear fitting of the binned samples.}
\label{results_NTD}
\fontsize{8pt}{8pt}\selectfont
\begin{tabular}{ccccccccc}

\hline
\multirow{2}{*}{\textbf{Sample}} & \multicolumn{3}{c}{\textbf{Line-Continuum}} & \multicolumn{3}{c}{\textbf{Baldwin Effect}} & \multirow{2}{*}{$N_{bins}$} & \multirow{2}{*}{$p_{bin}$} \\
\cline{2-7}
 & A & B & p-value & $\alpha$ & $\beta$ & p-value & & \\
 (1) & (2) & (3) & (4) & (5) & (6) & (7) & (8) & (9) \\
\hline
RQ $1<NTD<2$ & $4.622 \pm 0.065$ & $0.855 \pm 0.014$ & 0.0$^a$ & $7.962 \pm 0.071$ & $-0.143 \pm 0.013$ & 0.0$^a$ & 145 & 145 \\
RQ $NTD>2$ & $5.298 \pm 0.235$ & $0.836 \pm 0.026$ & 0.0$^a$ & $8.823 \pm 0.216$ & $-0.166 \pm 0.024$ & 0.0$^a$ & 43 & 42 \\
RQ $NTD<1$ & $8.348 \pm 0.150$ & $0.778 \pm 0.020$ & 0.0$^a$ & $11.383 \pm 0.154$ & $-0.213 \pm 0.020$ & 0.0$^a$ & 133 & 133 \\
FSRQ Disk-Dominated & $5.253 \pm 0.723$ & $0.841 \pm 0.066$ & $1.29 \times 10^{-5}$ & $9.412 \pm 0.618$ & $-0.174 \pm 0.037$ & $6.8 \times 10^{-4}$ & 12 & 12 \\
FSRQ Jet-Dominated$^b$ & $6.010 \pm 1.445$ & $0.818 \pm 0.178$ & $4.94 \times 10^{-4}$ & $8.067 \pm 2.004$ & $-0.152 \pm 0.129$ & $1.1 \times 10^{-2}$ & 7 & 7 \\
FSRQ $NTD<1$ & $11.965 \pm 1.483$ & $0.701 \pm 0.066$ & $1.79 \times 10^{-3}$ & $14.456 \pm 1.193$ & $-0.279 \pm 0.052$ & $2.18 \times 10^{-3}$ & 16 & 15 \\
\hline

\end{tabular}

\smallskip
\begin{minipage}{\textwidth}
\fontsize{8pt}{8pt}\selectfont
\textit{Note}: For the relationships $\log{(L_{MgII})}=A + B \times \log{(\lambda L_{3000})}$ the parameters are listed in columns 2-4, and for $\log{(EW_{MgII})}= \alpha + \beta \times \log{(\lambda L_{3000})}$ see columns 5-7. The number of bins ($N_{bins}$) is stated in column 8, and the points per bin ($p_{bin}$) are listed in column 9.\\
(a) To machine accuracy.\\
(b) It is important to note, the authors consider that the number of bins, and points per bin, may be too low to be a representative sample for the fit (and therefore, the relationship) to be considered for any conclusions.
\end{minipage}

\end{table*}

\begin{table*}
  \centering
  \makebox[0.75\textwidth]{ 
    \begin{threeparttable}
      \caption{Pearson correlation coefficients and p-values for the linear fits described in Table~\ref{results_NTD}.}
      \label{correlations_NTD}
      \fontsize{8pt}{8pt}\selectfont
      \begin{tabular}{ccccc}
        \hline
        \multirow{2}{*}{\textbf{Sample}} & \multicolumn{2}{c}{\textbf{Line-Continuum}} & \multicolumn{2}{c}{\textbf{Baldwin Effect}} \\
        \cline{2-5}
        & Pearson Coefficient & p-value & Pearson Coefficient & p-value \\
        \hline
        RQ $1<NTD<2$ & 1.0$^a$ & $1.01 \times 10^{-245}$ & -0.992 & $5.67 \times 10^{-129}$ \\
        RQ $NTD>2$ & 0.999 & $1.37 \times 10^{-61}$ & -0.985 & $5.28 \times 10^{-33}$ \\
        RQ $NTD<1$ & 0.999 & $1.81 \times 10^{-177}$ & -0.983 & $1.18 \times 10^{-97}$ \\
        FSRQ Disk-Dominated & 0.999 & $1.99 \times 10^{-14}$ & -0.975 & $7.77 \times 10^{-8}$ \\
        FSRQ Jet-Dominated$^b$ & 0.996 & $1.41 \times 10^{-6}$ & -0.924 & $2.96 \times 10^{-3}$ \\
        FSRQ $NTD<1$ & 0.991 & $1.80 \times 10^{-13}$ & -0.951 & $1.51 \times 10^{-8}$ \\
        \hline
      \end{tabular}
      
      \begin{tablenotes}
        \item(a) To machine accuracy.
        \item(b) As mentioned in Table~\ref{results_NTD}, the authors consider that the number of bins, and points per bin, may be too low for the fit (and the relationship) to be considered for any conclusions.
      \end{tablenotes}
    \end{threeparttable}
  } 
\end{table*}

A more detailed discussion regarding the NTD-separated samples can be found in Section~\ref{sec:Discussion}.

\section{The origin of the Baldwin effect} \label{sec:BE_origin}

As mentioned in Section~\ref{introduction}, the driving physical mechanism behind the BE has been the subject of study for decades. In this work, we test the hypothesis proposed by \cite{PatinoAlvarez2016}, which suggests that the BE is a natural consequence of the relationship between emission line luminosity and continuum luminosity.

This hypothesis arises from the fact that the BE’s mathematical expression, $\log{(EW_{MgII})}= \alpha + \beta \times \log{(\lambda L_{3000})}$, can be derived from the line-continuum relation, $\log{(L_{MgII})}=A + B \times \log{(\lambda L_{3000})}$, through simple algebra, under the assumption that $EW_{MgII} = L_{MgII} / L_{3000}$. For the data used in this work, this assumption holds true within the uncertainties for both the FSRQ and RQ sources (see Appendix~\ref{EW_comparison}). During this variable transformation, an important observable emerges: the relationship between the slope of the BE and the slope of the line- continuum relation, i.e., $\beta = B - 1$ or $B - \beta = 1$. For all the relations we calculated in this paper, we obtained the difference between these slopes, and the results are summarized in Table~\ref{slope_differences}.

\begin{table}
\centering
\caption{Differences between the slopes of the line luminosity - continuum luminosity relation, and the Baldwin effect.}
\label{slope_differences}
\begin{tabular}{cc}

\hline
\textbf{Sample} & \textbf{Slope Difference} \\
\hline
RQ Full Sample & $0.995 \pm 0.035$ \\

RQ $1<NTD<2$ & $0.997 \pm 0.018$ \\

RQ $NTD>2$ & $1.001 \pm 0.036$ \\

RQ $NTD<1$ & $0.992 \pm 0.028$ \\

FSRQ Full Sample & $0.978 \pm 0.126$ \\

FSRQ Disk-Dominated & $1.015 \pm 0.076$ \\

FSRQ Jet-Dominated$^b$ & $0.970 \pm 0.220$ \\

FSRQ $NTD<1$ & $0.980 \pm 0.084$ \\
\hline

\end{tabular}
\begin{threeparttable}
\begin{tablenotes}
\item[b] As noted in Table~\ref{results_NTD}, the authors suggest that the number of bins and points per bin may be insufficient for the fit (and the relationship) to support any definitive conclusions.
\end{tablenotes}
\end{threeparttable}
\end{table}

\section{Discussion} \label{sec:Discussion}

\subsection{NTD-separated samples}

As we did for the full samples, we tested whether there was a statistically significant difference between the subsamples of RQ and FSRQ sources. To do this, we applied a Welch unpaired t-test to the slopes and their uncertainties.

For the line-continuum relationship in the disk-dominated samples, the p-value is 0.4786, indicating that there is no statistically significant difference between the slope of the RQ sample and that of the FSRQ sample.

For the jet-dominated samples, it is important to note that the FSRQ subsample only contains 7 bins, which we consider too small to be representative. As a result, we do not regard the comparisons made with this subsample as significant, though we calculated them for completeness. We found a p-value of 0.7984 for the slopes of the line-continuum relationship, indicating that the difference is not statistically significant. Similarly, for the BE, the p-value is 0.7843, suggesting no statistically significant difference.

For the subsamples with NTD values lower than one, i.e., the objects where we suspect the accretion disk is not the only ionization source for the broad-line material, we found a p-value of $3.2 \times 10^{-4}$ when comparing the slopes of the line-continuum relation, indicating a statistically significant difference. For the BE, the analysis yielded a p-value of $1.5 \times 10^{-4}$, also indicating a statistically significant difference.

We also compared the relations obtained for the full RQ sample and the disk-dominated FSRQ subsample to test whether FSRQs in a low-activity state show a significant difference from RQ sources. The p-values obtained were 0.45 for the line-continuum relation and 0.65 for the BE, suggesting that there is no significant difference in the BLR emission between RQ sources and FSRQ-type blazars when the jet is in a low-activity state.

In contrast, when comparing the full RQ sample with the FSRQ sample having $NTD < 1$, the slopes show a statistically significant difference, with p-values of $1.8 \times 10^{-6}$ for the line-continuum relation and $4.8 \times 10^{-7}$ for the BE. This suggests that when the accretion disk is not the sole ionization source in FSRQs, the BLR emission is intrinsically different from that of RQ sources.

We also analyzed the differences in the NTD distributions using a Kolmogorov-Smirnov (K-S) test. The K-S test results for both the full RQ and full FSRQ samples show a statistically significant difference, which is consistent with the results obtained from comparing the slopes. For the disk-dominated samples and those with $NTD<1$, the K-S test yields the same conclusion as the slope comparisons. The only differing result comes from the jet-dominated samples, but as previously mentioned, we consider the slope from only 7 bins to be insufficiently representative of the population.

\subsection{Origin of the Baldwin effect}

As mentioned in Section~\ref{introduction}, numerous studies over the years have explored the driving mechanism behind the BE. These studies have considered various hypotheses, including the possibility that the ionizing continuum softens as luminosity increases, as well as factors such as the continuum shape, the metallicity of the gas, and the Eddington ratio, among others.

In Section~\ref{sec:BE_origin}, we discussed testing the hypothesis proposed by \cite{PatinoAlvarez2016}, which suggests that the BE is a mathematical consequence of the relationship between continuum luminosity and emission line luminosity. The key observable in the 'toy model' is that the difference between the slope of this relation and the slope of the BE is equal to one. It is worth noting that this was the case for the data in \cite{PatinoAlvarez2016}, but due to the smaller sample sizes, the authors chose not to emphasize this point in their paper.

As shown in the results of Table~\ref{slope_differences}, all slope differences are consistent with one. This suggests that the BE is indeed a consequence of the line luminosity - continuum luminosity relation for Mg II $\lambda 2798$ \AA\ and the 3000 \AA\ continuum. Therefore, no additional physical mechanism needs to be invoked to explain the existence of the BE.

\section{Results and conclusions} \label{sec:conclusions}

In this study, we analyzed the Mg II $\lambda$2798 \AA\ emission line and the 3000 \AA\ continuum emission for a sample of 40,685 radio-quiet quasars and 441 FSRQ-type blazars. We investigated the relationship between the emission line luminosity and continuum luminosity, as well as the Baldwin Effect and its potential origins. Below, we summarize our key findings and conclusions:

\begin{enumerate}

\item We revisited the relationship between the Mg II $\lambda$2798 \AA\ emission line and the 3000 \AA\ continuum, considering the natural dispersion of data due to AGN variability. After excluding over 3000 radio-loud sources and comparing our results with those of S11, we present a new relation. We applied binning to minimize variability effects, and the p-value of the fit suggests that the new relation accurately describes the data.

\item We found a statistically significant difference between the slopes of the line luminosity–continuum luminosity relation for the RQ and FSRQ samples. A similar difference was also observed for the Baldwin Effect. This suggests two possible explanations: (a) an intrinsic and systematic difference between the accretion disk spectrum of RQ AGNs and FSRQs, or (b) the continuum emission from the jet contributes to the ionization of Broad Line Region (BLR) material (either from the canonical BLR or an extended region) in FSRQ-type blazars.

\item The Non-Thermal Dominance (NTD) parameter shows that 43.8\% of the radio-quiet sample and 55.5\% of the blazar sample have NTD values lower than one. This indicates that, for blazars, the accretion disk is not the only source of ionization. On the other hand, for radio-quiet quasars we interpret this as a signature of several physical mechanisms: anomalies in the BLR structure (such as outflow or inflows), time lags between continuum and line variations, and the suppression of the UV continuum by a strong corona that diverts accretion power

\item For the Mg II $\lambda$2798 \AA\ emission line and the 3000 \AA\ continuum emission, we conclude that the Baldwin Effect is a natural consequence of the relationship between emission line luminosity and continuum luminosity. As a result, no additional physical mechanism is required to explain the existence of the Baldwin Effect.

\end{enumerate}

In upcoming papers, we aim to extend this analysis to the H$\beta$ emission line with the 5100 \AA\ continuum, as well as to the C IV $\lambda$1549 \AA\ emission line with the 1350 \AA\ continuum. Studying those additional lines could offer further insights into the emission-line and continuum relationships across different AGN types.

\begin{acknowledgements}
We thank the anonymous referee for their helpful comments and suggestion, that improved the quality of the paper. J.U.G.-G. and D.E.M.-W. acknowledge support from the the CONAHCYT (Consejo Nacional de Humanidades, Ciencia y Tecnolog\'ia) program for Ph.D. and M.Sc. studies, respectively. Additionally, this work was supported by the Max Plank Institute for Radioastronomy (MPIfR) - Mexico Max Planck Partner Group led by V.M.P.-A. ICG acknowledge financial support from DGAPA-UNAM grant IN-119123 and CONAHCYT grant CF-2023-G-100. This research was supported by the YSU, in the frames of the internal grant.

Funding for the Sloan Digital Sky Survey IV has been provided by the Alfred P. Sloan Foundation, the U.S. Department of Energy Office of Science, and the Participating Institutions. 

SDSS-IV acknowledges support and resources from the Center for High Performance Computing  at the University of Utah. The SDSS website is www.sdss4.org.

SDSS-IV is managed by the Astrophysical Research Consortium for the Participating Institutions of the SDSS Collaboration including the Brazilian Participation Group, the Carnegie Institution for Science, Carnegie Mellon University, Center for Astrophysics | Harvard \& Smithsonian, the Chilean Participation Group, the French Participation Group, Instituto de Astrof\'isica de Canarias, The Johns Hopkins University, Kavli Institute for the Physics and Mathematics of the Universe (IPMU) / University of Tokyo, the Korean Participation Group, Lawrence Berkeley National Laboratory, Leibniz Institut f\"ur Astrophysik Potsdam (AIP),  Max-Planck-Institut f\"ur Astronomie (MPIA Heidelberg), Max-Planck-Institut f\"ur Astrophysik (MPA Garching), Max-Planck-Institut f\"ur Extraterrestrische Physik (MPE), National Astronomical Observatories of China, New Mexico State University, New York University, University of Notre Dame, Observat\'ario Nacional / MCTI, The Ohio State University, Pennsylvania State University, Shanghai Astronomical Observatory, United Kingdom Participation Group, Universidad Nacional Aut\'onoma de M\'exico, University of Arizona, University of Colorado Boulder, University of Oxford, University of Portsmouth, University of Utah, University of Virginia, University of Washington, University of Wisconsin, Vanderbilt University, and Yale University.
\end{acknowledgements}

\bibliographystyle{aa} 
\bibliography{aa55413-25}

\begin{appendix} 

\section{Looking for non-blazars in the Roma BZCAT sample}
\label{non-blazars}

As noted in subsection~\ref{fsrq-sample}, we cannot be entirely certain that all sources in the 5th Roma Blazar Catalog (5BZCAT) are bona fide blazars. For instance, \cite{Xie2024} found that 4.5\% of the sources in the Roma-BZCAT were misclassified as blazars, indicating they are likely contaminants (i.e., non-blazars). Additionally, \cite{Kharb2010} examined a sample of 135 radio-loud active galactic nuclei (confirmed blazars) from the MOJAVE project and discovered that approximately 93\% of the sample displayed extended radio structures, while the remainder showed no such structures.

This prompted us to search for radio images of our FSRQ sample in order to assess potential contamination by non-blazar sources. We gathered 1.4 GHz radio images from the public database of the Faint Images of the Radio Sky at Twenty-cm (FIRST\footnote{\url{https://sundog.stsci.edu/cgi-bin/searchfirst}}, with an angular resolution of 5 arcseconds) for 90\% of the sources, and for the full sample, we used images from the NRAO VLA Sky Survey (NVSS\footnote{\url{https://www.cv.nrao.edu/nvss/findFITS.shtml}}, with an angular resolution of 45 arcseconds). Additionally, we obtained 144 MHz images for 64\% of the sources from the LOw-Frequency ARray (LOFAR\footnote{\url{https://lofar-surveys.org/dr2_release.html}}, with an angular resolution of 6 arcseconds), as this lower frequency helps trace older electrons and may provide insights into the more distant jet structures.

Blazars have jets that are oriented within a narrow angle of our line of sight, so the typical radio image of a blazar should either show a compact point spread function (PSF)-like source or a compact source with a one-sided jet, as the emission from the counter-jet is expected to be collimated away from us. Consequently, the primary criterion for suspecting that a source may not be a blazar is if the radio image exhibits a two-sided morphology, resembling that of a radio galaxy.

For our sample of 441 FSRQs, we identified 225 sources exhibiting a compact structure, 91 sources displaying a one-sided jet, and 95 sources showing a two-sided jet morphology. Notably, whenever we found a compact source in the 144 MHz images, the same morphology was also present in the 1.4 GHz images. However, the reverse was not always true—some sources exhibited a compact morphology at 1.4 GHz, but the 144 MHz images revealed a clear two-sided jet. Additionally, due to its low resolution, NVSS was not useful for distinguishing the morphology of these sources.

For the 95 sources with a two-sided jet morphology, we conducted a more detailed analysis to determine whether they are blazars. It is worth noting that previous studies, such as \cite{HernandezGarcia2023}, have shown that a source with a radio galaxy morphology can still have a blazar-like core. Thus, we decided to further analyze these sources to discern their true nature.

As discussed earlier, \cite{Kharb2010} conducted an analysis of MOJAVE VLBA data and VLA images, measuring the alignment of jets on both parsec and kiloparsec scales. They found that approximately 30\% of the jets were misaligned by more than 90 degrees, and they defined such cases as significantly misaligned. They argued that this misalignment suggests that while the kpc-scale jet may resemble a radio galaxy morphology, the inner jet could behave differently, allowing for the possibility that the AGN could have a blazar-like core. Furthermore, they pointed out that misalignments in parsec-scale jets might be caused by intrinsic factors within the galaxy’s active core or its environment, such as the presence of binary black holes or black hole mergers, which could cause changes in the direction of jets in some blazars.

Based on this understanding, we proceeded to measure the jet angles on both scales for the 95 sources showing a two-sided jet morphology. The angle was measured by setting the west horizontal as the zero angle, tracing the direction from the core to the brightest lobe (in the case of the FIRST and LOFAR images), and from the core to the projected direction of the jet in the VLBI images. For verification, we compared our results with those of \cite{Plavin2022}, who developed an automated method for measuring jet directions at parsec scales across 9220 AGNs using frequencies ranging from 1.4 to 86 GHz. Our measurements for the same sources were consistent with theirs. The VLBI images were obtained from the Astrogeo VLBI FITS image database\footnote{\url{http://astrogeo.org/vlbi_images/}}.

It is noteworthy that several sources classified as compact by \cite{Xie2024} are identified as extended in our 144 MHz images. In an upcoming publication, we will present a comprehensive catalog of blazars with optical spectra from the Sloan Digital Sky Survey, incorporating the information discussed here, along with additional multi-frequency data and a thorough statistical analysis.

Since jet angles alone may not conclusively determine whether a source is a blazar, we sought other criteria to help distinguish between blazars and radio galaxies. One key characteristic of blazars is their high variability on short time scales. This led us to explore databases focused on AGN or variable object monitoring. The most comprehensive database we found is the Zwicky Transient Facility \citep[ZTF;][]{Bellm2019,Masci2019}, which conducts a repetitive survey with photometric observations in the g, r, and i optical bands. Given the limited data in the i band, we primarily used the g and r bands.

For the light curves, we classified a source as a blazar if it exhibited fast flaring events, typical of the time contraction due to Doppler boosting \citep{Sher1968}. Such flaring is a strong indicator that the jet emitting optical synchrotron radiation—expected to originate from the inner jet a few parsecs from the black hole— is pointed close to our line of sight.

\begin{figure*}[htbp]
\begin{center}
\includegraphics[width=0.68\textwidth]{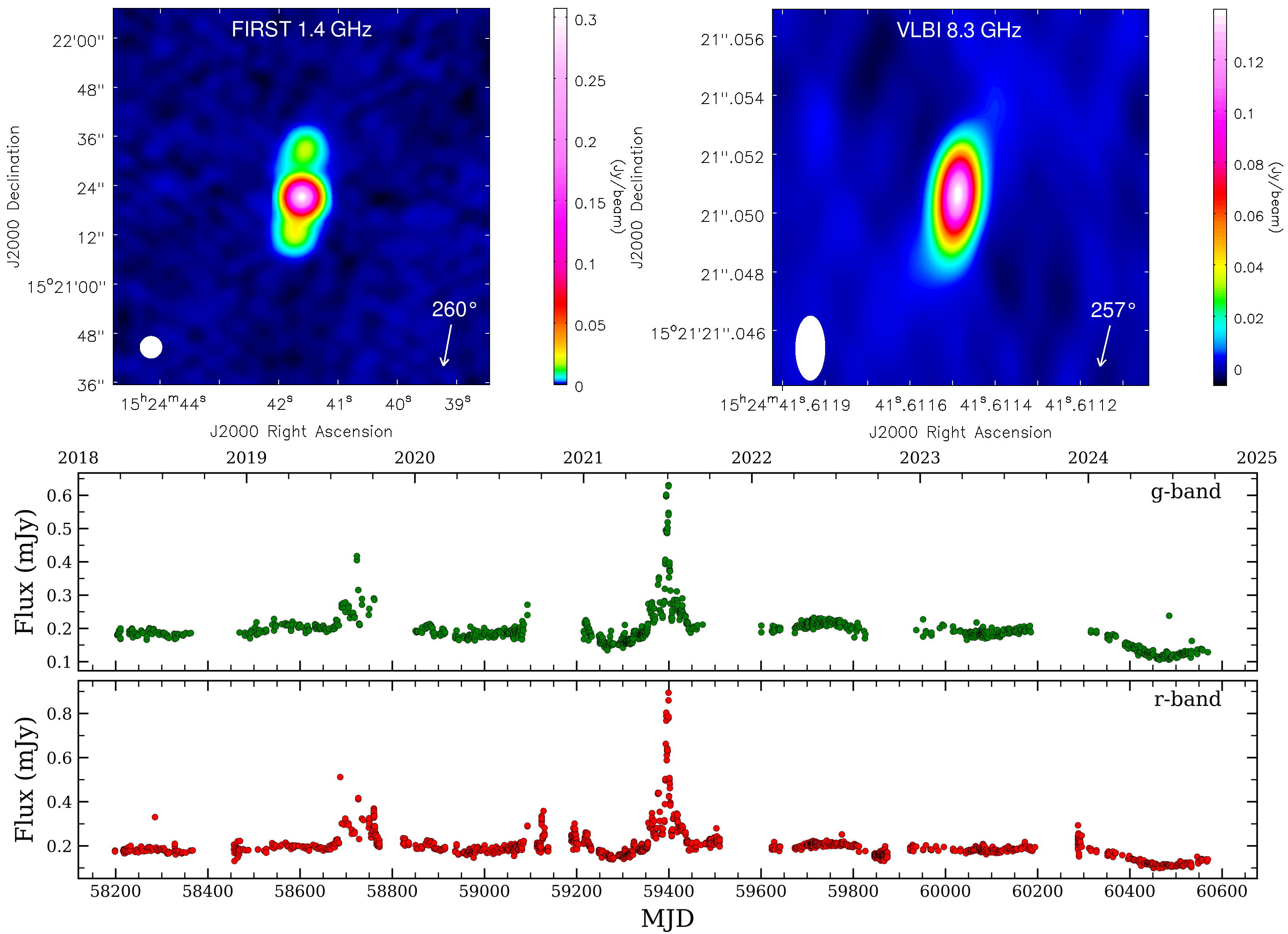}
\caption{Source 5BZQ J1524+1521. Top left: 1.4 GHz image from FIRST, showing the jet inclination at the kpc scale. Top right: VLBI image at 8.3 GHz, showing the jet inclination at the pc scale. Middle: Optical light curve in the g band, retrieved from the ZTF database. Bottom: Optical light curve in the r band, retrieved from the ZTF database.}
\label{Jet-angle-lc-1}
\end{center}
\end{figure*}

\begin{figure*}[htbp]
\begin{center}
\includegraphics[width=0.68\textwidth]{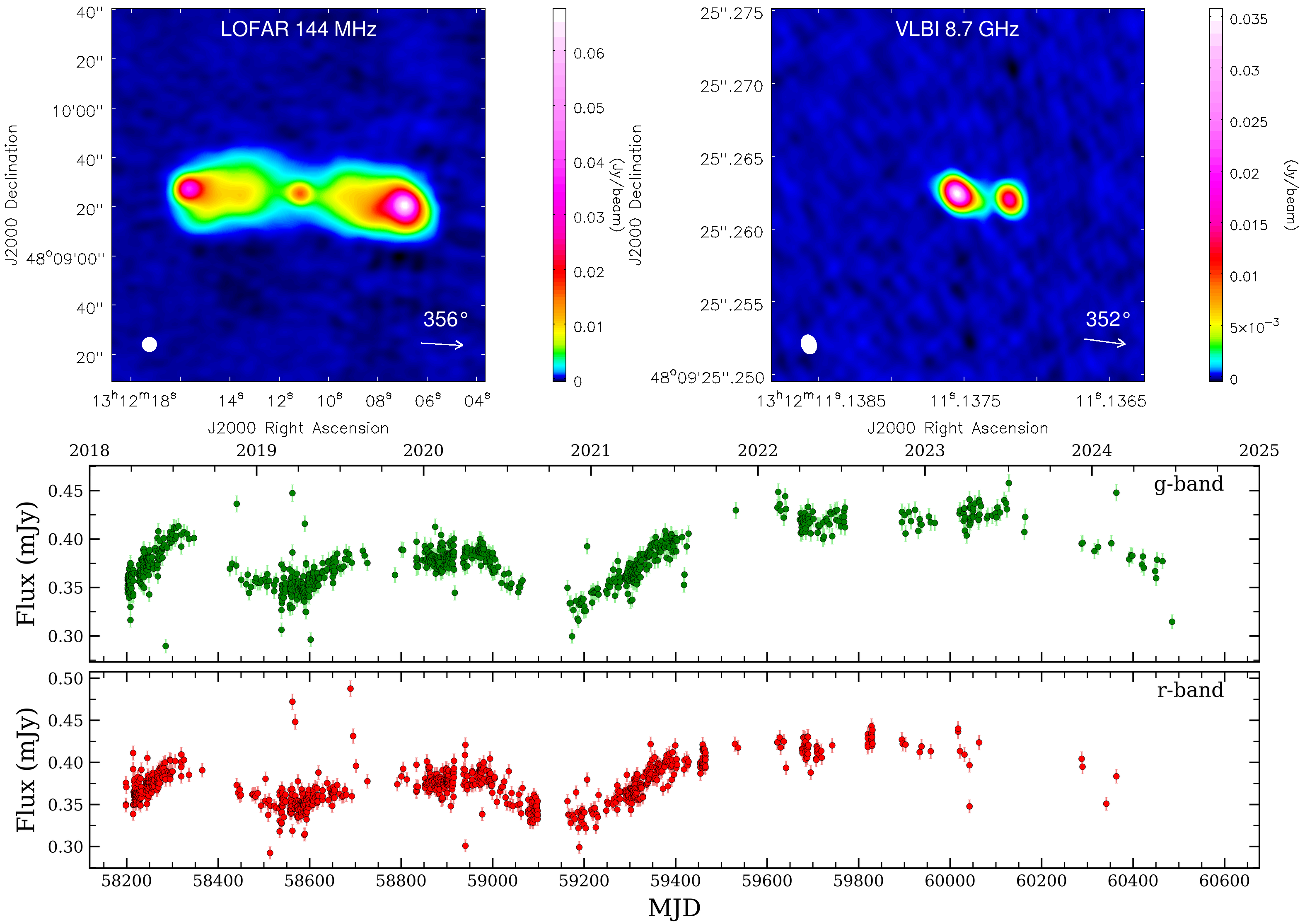}
\caption{Source 5BZQ J1312+4809. The panels are the same as in figure~\ref{Jet-angle-lc-1}.}
\label{Jet-angle-lc-2}
\end{center}
\end{figure*}

\begin{figure*}[htbp]
\begin{center}
\includegraphics[width=0.68\textwidth]{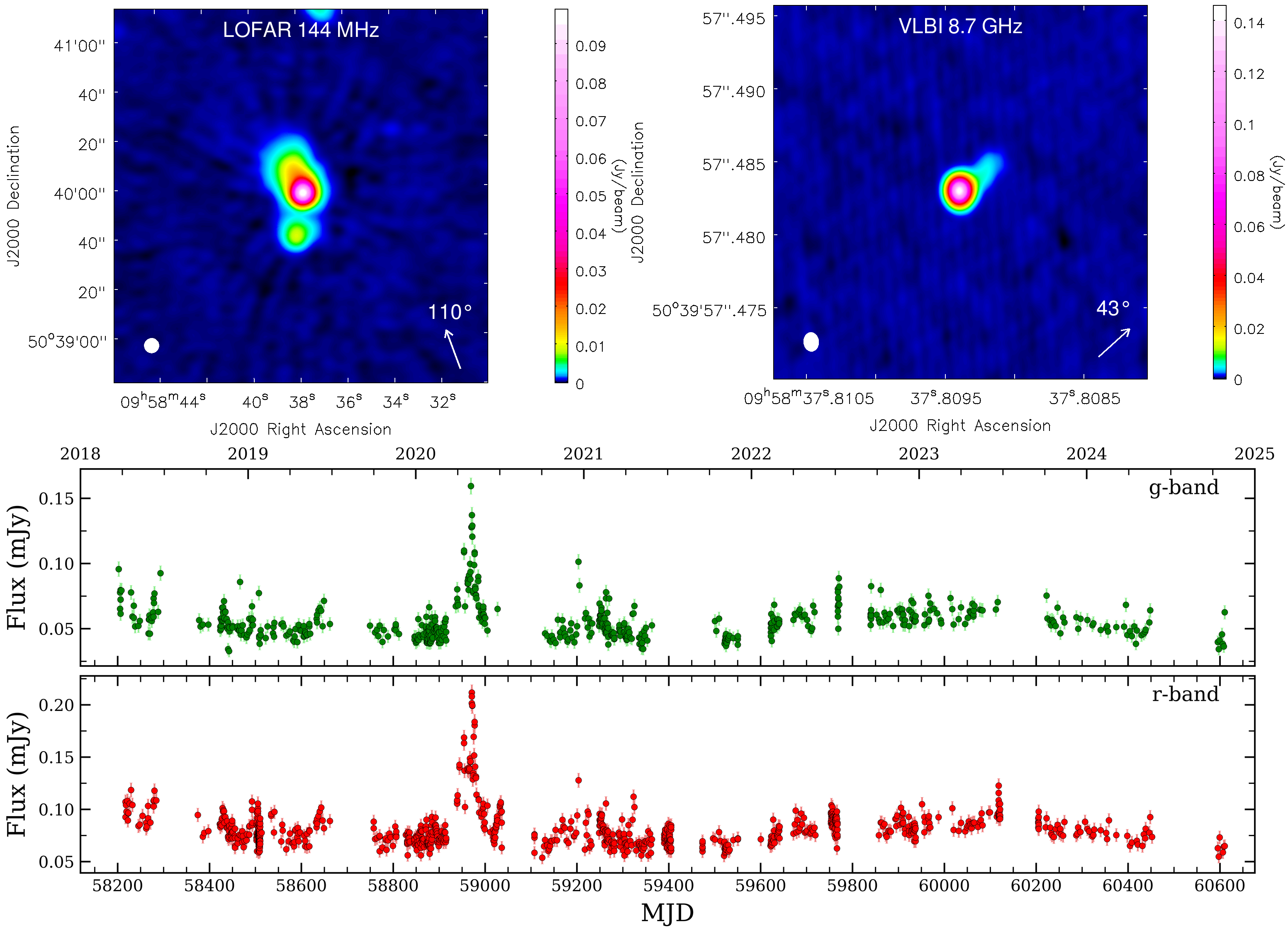}
\caption{Source 5BZQ J0958+5039. The panels are the same as in figure~\ref{Jet-angle-lc-1}.}
\label{Jet-angle-lc-3}
\end{center}
\end{figure*}

\begin{figure*}[htbp]
\begin{center}
\includegraphics[width=0.68\textwidth]{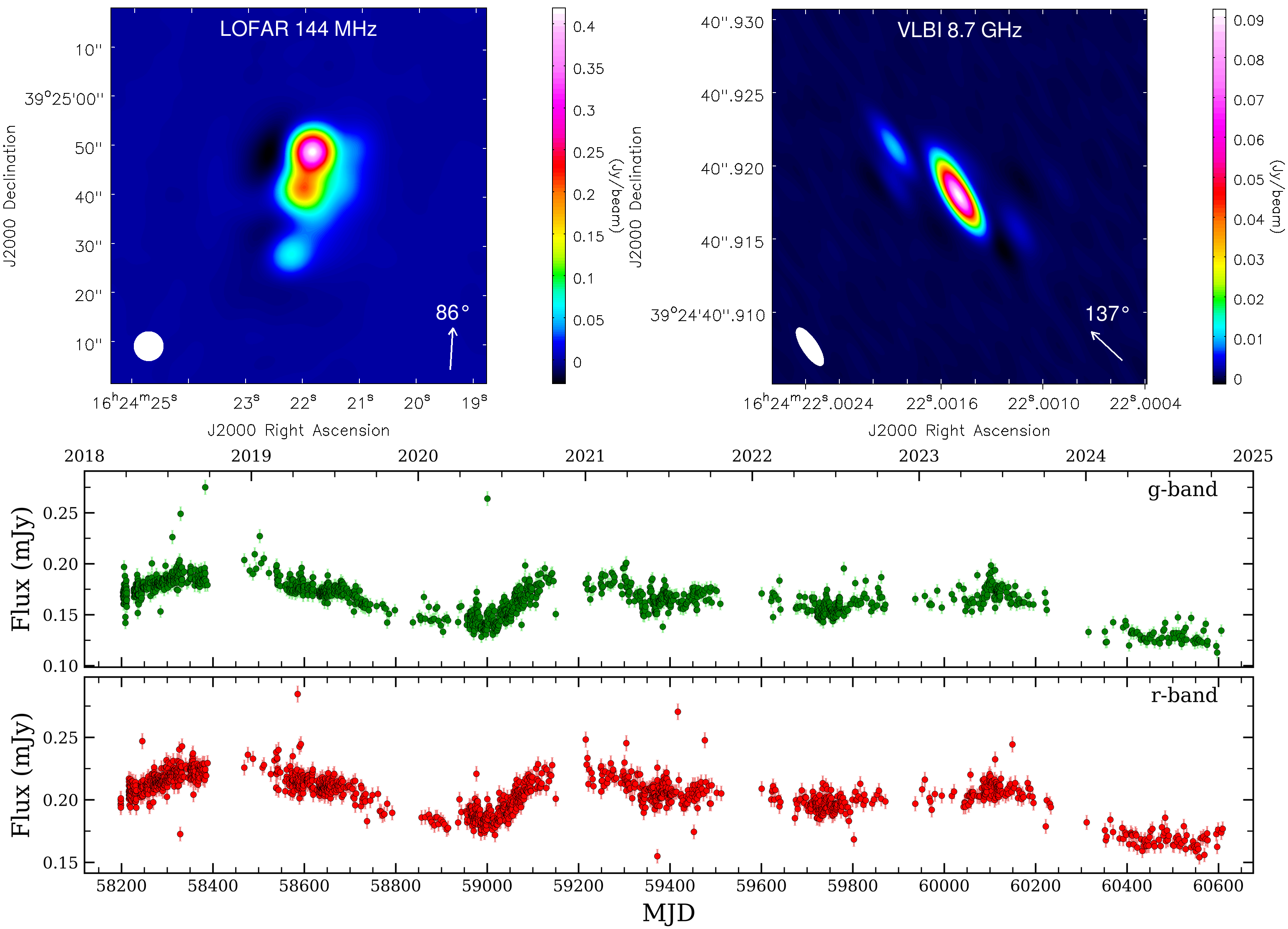}
\caption{Source 5BZQ J1624+3924. The panels are the same as in figure~\ref{Jet-angle-lc-1}.}
\label{Jet-angle-lc-4}
\end{center}
\end{figure*}

Based on these criteria, we found 72 sources (16\% of the sample) for which we cannot definitively confirm their blazar status. In Figures~\ref{Jet-angle-lc-1}-\ref{Jet-angle-lc-4}, we present four examples illustrating the following scenarios: i) the jet is aligned on both parsec and kiloparsec scales, and the light curve shows blazar-like behavior (Figure~\ref{Jet-angle-lc-1}); ii) the jet is aligned, but the light curve does not exhibit flaring behavior (Figure~\ref{Jet-angle-lc-2}); iii) the jet is misaligned, yet the light curve shows blazar-like behavior (Figure~\ref{Jet-angle-lc-3}); and iv) the jet is misaligned, and the light curve does not show flaring behavior (Figure~\ref{Jet-angle-lc-4}).

This leaves us with a "clean" sample of 369 FSRQs, for which we decided to repeat the correlation analysis and perform the fits for the Line Luminosity - Continuum Luminosity relation and the Baldwin Effect, excluding the 72 sources. As before, we conducted this analysis for both the full sample and the subsamples separated by NTD. The results are summarized in Table~\ref{clean_sample_results}. For the fits that accurately represent the data (i.e., p-value $<0.05$), we performed a Welch unpaired t-test to compare the slopes obtained for the "clean" sample of 369 sources with those from the full sample of 441 FSRQs. In every case, the conclusion was that there is no statistically significant difference between the slopes of the relationships, meaning they are equivalent. Therefore, in order to maximize the number of data points, we decided to proceed with the full sample.

\begin{table*}
\centering
\caption{Parameters obtained from the linear fitting of the "clean" sample of FSRQ.}
\label{clean_sample_results}
\fontsize{7pt}{7pt}\selectfont
\begin{tabular}{ccccccccc}

\hline
\multirow{2}{*}{\textbf{Sample}} & \multicolumn{3}{c}{\textbf{Line-continuum}} & \multicolumn{3}{c}{\textbf{Baldwin effect}} & \multirow{2}{*}{$N_{bins}$} & \multirow{2}{*}{$p_{bin}$} \\
\cline{2-7}
 & A & B & p-value & $\alpha$ & $\beta$ & p-value & & \\
\hline
FSRQ (Clean) Full Sample & $11.107 \pm 0.991$ & $0.717 \pm 0.089$ & $1.46 \times 10^{-8}$ & $13.773 \pm 0.818$ & $-0.266 \pm 0.083$ & $1.98 \times 10^{-9}$ & 19 & 19 \\

FSRQ (Clean) Disk-Dominated & $6.556 \pm 0.672$ & $0.813 \pm 0.059$ & $3.2 \times 10^{-4}$ & $10.172 \pm 1.000$ & $-0.191 \pm 0.037$ & 0.0675$^a$ & 9 & 9 \\

FSRQ (Clean) Jet-Dominated & $5.148 \pm 3.989$ & $0.836 \pm 0.146$ & 0.162$^a$ & $8.752 \pm 3.089$ & $-0.167 \pm 0.129$ & 0.107$^a$ & 6 & 6 \\

FSRQ (Clean) $NTD<1$ & $10.842 \pm 1.102$ & $0.726 \pm 0.069$ & $4.35 \times 10^{-4}$ & $14.503 \pm 0.687$ & $-0.279 \pm 0.041$ & $6.84 \times 10^{-4}$ & 12 & 12 \\
\hline

\end{tabular}

\smallskip
\begin{minipage}{\textwidth}
\fontsize{7pt}{7pt}\selectfont
\textit{Note}: For the relationships $\log{(L_{MgII})}=A + B \times \log{(\lambda L_{3000})}$ the parameters are listed in columns 2-4, and for $\log{(EW_{MgII})}= \alpha + \beta \times \log{(\lambda L_{3000})}$, see columns 5-7. The number of bins ($N_{bins}$) is stated in column 8, and the points per bin ($p_{bin}$) are listed in column 9.\\
(a) These p-values indicate that the linear fit does not accurately represents the data. It is also possible that the low number of bins and points per bin, prevents us from having a representative sample.
\end{minipage}

\end{table*}

\FloatBarrier

\section{Comparing of the NTD distributions}
\label{NTD_comparison}

The main challenge in directly comparing the Non-Thermal Dominance (NTD) values between the Radio-Quiet (RQ) and FSRQ samples, lies in the substantial difference in their sizes—nearly two orders of magnitude—which poses problems for traditional statistical tests \citep{Hollander2014} such as the Kolmogorov–Smirnov (K–S) test, the Anderson–Darling (A–D) test, and the Mann–Whitney U (MWU) test. The reasons are as follows:

\begin{itemize}

\item Kolmogorov–Smirnov test:
When sample sizes are highly unbalanced, the K–S test can detect statistically significant but practically negligible differences. Because the empirical cumulative distribution function (ECDF) of the large sample is estimated with high precision, even a minute deviation can appear statistically significant—a phenomenon commonly referred to as the over-power problem.
\\ 
\item Anderson–Darling test:
The A–D test suffers from a similar over-power issue, often to a greater extent due to its increased sensitivity in the distribution tails.
\\ 
\item Mann–Whitney U test:
The MWU test is based on ranking all observations together under the null hypothesis. When one sample is much larger, its ranks dominate the pool, effectively turning the test into a comparison of the smaller sample’s ranks against the “background” of the larger sample. This imbalance can reduce statistical power and hinder the detection of genuine location differences.

\end{itemize}

To address the large disparity in sample sizes (N$_{RQ}$ = 40,685; N$_{FSRQ}$ = 441), we employed bootstrap resampling to construct subsamples from the RQ sample that retained the same statistical properties but were of comparable size to the FSRQ sample. This approach allows fairer comparisons without the confounding effects of unequal sample sizes.

Although there is no strict limit, statistical literature and software documentation commonly suggest that the larger sample should not exceed 4–5 times the size of the smaller one. Following this principle, we tested subsample sizes ranging from 100 to 2,000 (in increments of 100). For each subsample size, we performed 10,000 bootstrap resamplings (without replacement) and identified the optimal subsample size based on a combination of five distributional similarity metrics:

\begin{itemize}

\item Maximum distance between empirical cumulative distribution functions \citep{Massey1951}.
\\ 
\item One-dimensional Wasserstein distance \citep{Peyre2019}.
\\ 
\item Standardized mean difference \citep[Cohen’s d;][]{Cohen1988}.
\\ 
\item Ratio of standard deviations.
\\ 
\item Absolute differences in skewness and kurtosis coefficients \citep{Sheskin2011}.

\end{itemize}

For each of the 10,000 iterations, the metrics for all candidate subsample sizes were evaluated, and the subsample with the closest distribution to the full RQ sample was recorded. Across all five metrics, the subsample size of n = 2,000 was most frequently identified as the best match.

From the 10,000 candidate subsamples of size n = 2,000, the 5 most representative subsamples were then selected by minimizing the same set of metrics (one sample for each metric). This procedure ensures that subsequent statistical comparisons between the FSRQ sample (n = 441) and the optimally selected RQ subsamples (n = 2,000) are not biased by differences in sample size, while preserving the distributional characteristics of the original large sample.

The exact same procedure was applied to the samples while separated by the three NTD regimes discussed in Section~\ref{sec:NTD}. For the disk-dominance regime ($1<NTD<2$), the ideal RQ subsample size is 720, to compare with the 144 FSRQ. In the jet-dominated regime ($NTD>2$), the best RQ subsample size is 250, to compare with the 52 FSRQ. Finally, in the $NTD<1$ regime, the analysis concluded that a RQ subsample size of 1250, is the ideal size to compare with the 245 FSRQ. As with the full sample, the 5 most representative subsamples were selected from the metrics results.

The final statistical comparison between the selected RQ subsamples and the FSRQ samples was conducted using the Kolmogorov–Smirnov test, the Anderson–Darling test, and the Mann–Whitney U test \citep{Hollander2014}. The results are shown in Table~\ref{tab:stat_results}.

\begin{table*}[htbp]
\centering
\caption{Statistical test results (p-values) comparing FSRQ and RQ NTD distributions for the full samples and different NTD regimes.}
\label{tab:stat_results}
\begin{tabular}{lccccc}
\hline
\textbf{Sample} & \textbf{ECDF} & \textbf{Wasserstein} & \textbf{Std. mean diff.} & \textbf{Stddev ratio} & \textbf{High moment diff.} \\
\hline
\textbf{Full Sample} & & & & & \\
\quad K-S & \(1.426 \times 10^{-9}\) & \(1.725 \times 10^{-9}\) & \(1.426 \times 10^{-9}\) & \(1.614 \times 10^{-9}\) & \(1.582 \times 10^{-9}\) \\
\quad A-D & \(<1 \times 10^{-4}\) & \(<1 \times 10^{-4}\) & \(<1 \times 10^{-4}\) & \(<1 \times 10^{-4}\) & \(<1 \times 10^{-4}\) \\
\quad MWU & \(2.960 \times 10^{-6}\) & \(2.470 \times 10^{-6}\) & \(2.923 \times 10^{-6}\) & \(1.201 \times 10^{-6}\) & \(4.716 \times 10^{-5}\) \\
\hline
\textbf{Disk-Dominated} & & & & & \\
\quad K-S & 0.198 & 0.198 & 0.240 & 0.240 & 0.084 \\
\quad A-D & 0.259 & 0.358 & 0.269 & 0.343 & 0.139 \\
\quad MWU & 0.350 & 0.389 & 0.364 & 0.495 & 0.143 \\
\hline
\textbf{Jet-Dominated} & & & & & \\
\quad K-S & \(3.715 \times 10^{-11}\) & \(1.223 \times 10^{-10}\) & \(8.255 \times 10^{-11}\) & \(3.076 \times 10^{-12}\) & \(3.710 \times 10^{-11}\) \\
\quad A-D & \(<1 \times 10^{-4}\) & \(<1 \times 10^{-4}\) & \(<1 \times 10^{-4}\) & \(<1 \times 10^{-4}\) & \(<1 \times 10^{-4}\) \\
\quad MWU & \(1.577 \times 10^{-13}\) & \(1.197 \times 10^{-13}\) & \(1.197 \times 10^{-13}\) & \(3.061 \times 10^{-15}\) & \(1.706 \times 10^{-13}\) \\
\hline
\textbf{NTD < 1} & & & & & \\
\quad K-S & \(1.080 \times 10^{-8}\) & \(2.985 \times 10^{-8}\) & \(4.665 \times 10^{-9}\) & \(3.737 \times 10^{-9}\) & \(3.187 \times 10^{-9}\) \\
\quad A-D & \(<1 \times 10^{-4}\) & \(<1 \times 10^{-4}\) & \(<1 \times 10^{-4}\) & \(<1 \times 10^{-4}\) & \(<1 \times 10^{-4}\) \\
\quad MWU & \(5.888 \times 10^{-12}\) & \(1.096 \times 10^{-11}\) & \(7.598 \times 10^{-12}\) & \(2.358 \times 10^{-12}\) & \(3.566 \times 10^{-13}\) \\
\hline
\end{tabular}

\smallskip
\begin{minipage}{\textwidth}
\footnotesize
\textit{Note 1}: The comparisons are based on different metrics used to select the optimal bootstrap subsample: Empirical Cumulative Distribution Function (ECDF, column 2), Wasserstein distance (column 3), standardized mean difference (column 4), standard deviation ratio (column 5), and higher moment differences (column 6). p-values lower than 0.05 indicate statistically significant differences.\\
\textit{Note 2}: P-values reported as \(<1 \times 10^{-4}\) for the Anderson-Darling (A-D) test indicate that the calculated value is below the lower precision limit of the \texttt{scipy.stats.anderson\_ksamp} function in Python.
\end{minipage}
\end{table*}

In summary, all three statistical tests reached the same conclusion, regardless of which of the 5 subsamples was analyzed. No statistically significant difference was found in the NTD distribution, between the RQ AGN and the FSRQ, for the disk-dominance regime. On the other hand, significant differences were found when analyzing the full sample, the objects in the jet-dominated regime, and the sources with $NTD<1$.

\FloatBarrier

\section{Comparing equivalent width estimations}
\label{EW_comparison}

As mentioned in Section~\ref{sec:BE_origin}, \cite{PatinoAlvarez2016} propose that the Baldwin Effect is a natural consequence of the existence of the line luminosity - continuum luminosity relation. This assumption holds true when [$EW_{MgII} = L_{MgII} / L_{3000}$]. To verify this for the data used in this work, we calculated [$L_{MgII} / L_{3000}$] and compared the resulting values with the measured equivalent widths. The histograms of the differences are shown in Figure~\ref{EW_differences_hist}. For the FSRQ sample, we find that the mean of the differences is comparable to the uncertainties in the measured EWs. For the RQ sample, the mean of the differences is larger than the uncertainties reported by S11. However, as noted in Subsection~\ref{flux_measure}, we believe these uncertainties are likely underestimated. This leads us to the conclusion that the measured EWs and those estimated from [$EW_{MgII} = L_{MgII} / L_{3000}$] are consistent with each other within the uncertainties.

\begin{center}
\begin{minipage}{\textwidth}
\centering
\includegraphics[width=0.8\textwidth]{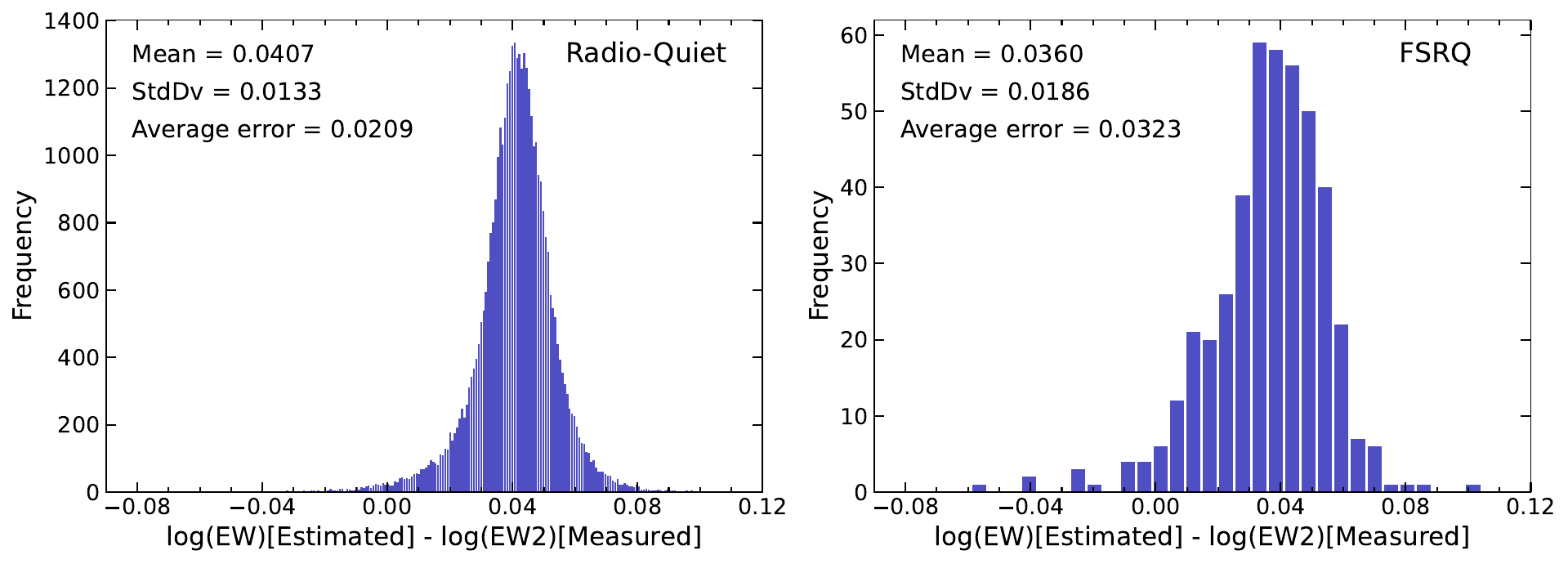}
\captionof{figure}{Differences between the EW estimated as [$L_{MgII} / L_{3000}$], and the measured values. Left panel: For the RQ sample. Right panel: For the FSRQ sample.}
\label{EW_differences_hist}
\end{minipage}
\end{center}

\end{appendix}
\end{document}